\documentclass[reqno]{article}
\usepackage{amssymb,amsmath,cite,delarray,epsfig,hhline}
\newcommand{\lsim} {\buildrel < \over {_\sim}}
\newcommand{\gsim} {\buildrel > \over {_\sim}}

\begin{document}

\title{\textbf{ Distinguishing Between Hierarchical and Lop-sided
SO(10) Models.}}
\author{Parul Rastogi\\
Department of Physics, University of Maryland, College Park, MD
20740 USA}\maketitle

\vskip.25in
\begin{abstract}
A comparative study of two predictive SO(10) models , namely the
BPW model (proposed by Babu, Pati and Wilczek) and the AB model
(proposed by Albright and Barr) is done based on their predictions
regarding CP and flavor violations. There is a significant
difference in the structure of the fermion mass-matrices in the
two models (which are hierarchical for the BPW case and lop-sided
for the AB model) which gives rise to different CP and flavor
violating effects. We include both SM and SUSY contributions to
these processes. Assuming flavor universality of SUSY breaking
parameters at a messenger scale $M^*\gsim M_{GUT}$, it has been
shown that renormalization group based post-GUT physics gives rise
to large CP and flavor violations. While these effects were
calculated for the BPW model recently, this is the first time (to
our knowledge) that post-GUT contributions have been included for
the AB model. The values of $\Delta m_K, \epsilon_K, \Delta
m_{B_d}$ and $S(B_d\to J/\psi K_S)$  are found, in both models, to
be close to SM predictions, in good agreement with data. Both
models predict that $S(B_d\to \phi K_S)$ should lie in the range
+0.65--0.74, close to the SM prediction and that the EDM of the
neutron $\approx (few\times 10^{-26})$e-cm, which should be
observed in upcoming experiments. The lepton sector brings out
marked differences between the two models. It is found that
$Br(\mu\to e\gamma)$ in the AB model is generically much larger
than that in the BPW model, being consistent with the experimental
limit only with a rather heavy SUSY spectrum with
$(m_o,~m_{1/2})\sim (1000,~1000)$ GeV. The BPW model, on the other
hand, is consistent with the SUSY spectrum being as light as
$(m_o,~m_{1/2})\sim (600,~300)$ GeV. Another distinction arises in
the prediction for the EDM of the electron. In the AB model $d_e$
should lie in the range $10^{-27}-10^{-28}$e-cm, and should be
observed by forthcoming experiments. The BPW model gives $d_e$ to
be typically 100 times lower than that in the AB case. Thus the
two models can be distinguished based on their predictions
regarding CP and flavor violating processes, and can be tested in
future experiments.
\end{abstract}
\newpage
{\large

\section{Introduction}

Grand unified theories \cite{JCPAS,PS,SU(5)} have found much
success in explaining (a) the quantum numbers of the members in a
family, (b) quantization of electric charge and (c) the meeting of
the gauge couplings at a scale $\sim 2\times 10^{16}$ GeV in the
context of supersymmetry \cite{GQW,DimWil}. In particular, it has
been argued \cite{JCPKeK} that the features of (d) neutrino
oscillations \cite{sk,sno}, (e) the likely need for baryogenesis
via leptogenesis \cite{Yanagida,PatiLepto}, and (f) the success of
certain mass relations like $m_b\approx m_\tau$ and
$m(\nu_\tau)_{\rm Dirac}\approx m_t$ at the unification scale,
suggest that the effective symmetry near the string/GUT scale in
4D should possess the symmetry SU(4)-color \cite{PS}. Thus, it
should be either SO(10) \cite{SO(10)} or minimally G(224) =
$SU(2)_L\times SU(2)_R\times SU(4)^c$ \cite{PS}. (For a detailed
review of the advantages and successes of G(224)/SO(10) symmetry,
see e.g. \cite{JCPKeK}.)

In recent years, several models based on supersymmetric SO(10) GUT
have emerged \cite{SO(10)GUT}. Two promising candidates have been
proposed which have much similarity in their Higgs structure and
yet important differences in the pattern of fermion mass-matrices.
One is by Albright and Barr (AB)\cite{AB} and the other by Babu,
Pati and Wilczek (BPW) \cite{BPW}. Both models use low-dimensional
Higgs multiplets (like $45_H, 16_H,\overline{16_H}$ and $10_H$) to
break SO(10) and generate fermion masses (see remarks later) as
opposed to large-dimensional ones (like $126, \overline{126}, 210$
and possibly 120). Both of these models work extremely well in
making predictions regarding the masses of quarks and leptons, the
CKM elements and neutrino masses and their mixings in good accord
with observations. Nevertheless there is a significant difference
between these two models in the structure of their fermion mass
matrices. In the BPW-model, the elements of the fermion
mass-matrices (constrained by a U(1)-flavor symmetry
\cite{Hall,JCPKeK,BPR}) are consistently family-hierarchical with
``33''$\gg$``23''$\sim$``32''$\gg$``22''$\gg$``12''$\sim$``21''$\gg$``11''
{\it etc.} By contrast, in the AB-model, the fermion mass-matrices
are lopsided with ``23''$\sim$``33'' in the down quark mass-matrix
and ``32''$\sim$``33'' in the charged lepton matrix. (The exact
structure of the fermion mass-matrices will be presented in Sec.
2.) This difference in the structure of the mass matrices leads to
two characteristically different explanations for the largeness of
the $\nu_\mu-\nu_\tau$ oscillation angle in the two models. For
the BPW model, both charged lepton and neutrino sectors give
moderately large contributions to this mixing which, as they show,
naturally add to give a nearly maximal $\sin^2
2\theta_{\nu_\mu-\nu_\tau}$, while simultaneously giving small
$V_{cb}$ as desired. The largeness of $\theta_{\nu_\mu-\nu_\tau}$,
together with the smallness of $V_{cb}$ (in the BPW model) turns
out in fact to be a consequence of (a) the group theory of
SO(10)/G(224) in the context of the minimal Higgs system, and (b)
the hierarchical pattern of the mass-matrices. For the lopsided AB
model, on the other hand, the large (maximal) $\nu_\mu-\nu_\tau$
oscillation angle comes almost entirely from the charged lepton
sector which has a ``32'' element comparable to the ``33''.

The original work of Babu, Pati and Wilczek, treated the entries
in the mass matrices to be real for simplicity, thereby ignoring
CP non-conservation. It was successfully extended to include CP
violation by allowing for phases in the mass matrices by Babu,
Pati and the author in Ref. \cite{BPR}.

The purpose of this paper is to do a comparative study between
certain testable predictions of  the AB model versus those of the
BPW model allowing for the extension of the latter as in Ref.
\cite{BPR}. We find that while both models give similar
predictions regarding fermion masses and mixings, they can be
sharply distinguished by lepton flavor violation, especially by
the rate of $\mu\to e\gamma$ and the edm of the electron.

We work in a scenario as in Refs. \cite{BPR} and \cite{LFV}, in
which flavor-universal soft SUSY breaking is transmitted to the
sparticles at a messenger-scale M$^{*}$, with M$_{GUT}< $
M$^{*}\le$ M$_{string}$ as in a mSUGRA model \cite{mSUGRA}.
Following the general analysis in Ref. \cite{BHS} it was pointed
out in Refs. \cite{BPR} and \cite{LFV} that in a SUSY-GUT model
with a high messenger scale as above, post-GUT physics involving
RG running from M$^*\to$ M$_{GUT}$ leads to dominant flavor and CP
violating effects. In the literature, however, post-GUT
contribution has invariably been omitted, except for Refs.
\cite{BPR} and \cite{LFV}, where it has been included only for the
BPW model. Lepton flavor violation in the AB model has been
studied so far by many authors by including the contribution
arising only through the RH neutrinos \cite{ABLFV}, without,
however, the inclusion of post-GUT contributions. I therefore make
a comparative study of the BPW and the AB models by including the
contributions arising from both post-GUT physics, as well as those
from the RH neutrinos through RG running below the GUT scale. For
the sake of comparison and completeness, we will include the
results obtained in Refs. \cite{BPR} and \cite{LFV} which deal
with CP and flavor violation in the BPW model.

To calculate the branching ratio of lepton flavor violating
processes we include contributions from three different sources:
(i) the sfermion mass-insertions, $\hat{\delta}^{ij}_{LL,RR}$,
arising from renormalization group (RG) running from M$^{*}$ to
M$_{GUT}\sim 2\times 10^{16}$ GeV, (ii) the mass-insertions
$(\delta^{ij}_{LL})^{RHN}$ arising from RG running from M$_{GUT}$
to the right handed neutrino mass scales M$_{R_i}$, and (iii) the
chirality-flipping mass-insertions $\delta^{ij}_{LR,RL}$ arising
from $A-$terms that are induced solely through RG running from
M$^{*}$ to M$_{GUT}$ involving SO(10) or G(224) gauginos in the
loop.

It was found in Ref. \cite{LFV}, that for the BPW-model,
contributions to the rate of $\mu\to e\gamma$  from sources (i)
and (iii) associated with post-GUT physics, were typically much
larger than that from source (ii) associated with the RH
neutrinos. For the AB-model, we find that the RH neutrino
contribution is strongly enhanced compared to that in the BPW
model; as a result all the three contributions to the amplitude of
$\mu\to e\gamma$ are comparable. Including all three
contributions, we find that for most of the SUSY parameter space,
the branching ratio for $\mu\to e\gamma$ calculated in the
AB-model is much larger than that in the BPW model and is in fact
excluded by the experimental upper bound unless $(m_o,\
m_{1/2})\gsim $ 1 TeV. Thus one main result of this paper is that,
with all three sources of lepton flavor violation included, the
process $\mu\to e\gamma$ can provide a clear distinction between
the BPW and the AB models. We also examine CP violation as well as
flavor violation in the quark sector, including that reflected by
electric dipole moments, in the AB model, and compare it with the
corresponding results for the BPW-model, obtained in \cite{BPR}.

In the following section the patterns of the fermion mass matrices
for the BPW and the AB models are presented.

\section{A brief description of the BPW and the AB models}

{\bf The Babu-Pati-Wilczek (BPW) model}

The Dirac mass matrices of the sectors $u, d, l$ and $\nu$
proposed in Ref. \cite{BPW} in the context of SO(10) or
G(224)-symmetry have the following structure:

\begin{eqnarray}
\label{eq:mat}
\begin{array}{cc}
M_u=\left[
\begin{array}{ccc}
0&\epsilon'&0\\-\epsilon'&\zeta_{22}^u&\sigma+\epsilon\\0&\sigma-\epsilon&1
\end{array}\right]{\cal M}_u^0;&
M_d=\left[
\begin{array}{ccc}
0&\eta'+\epsilon'&0\\
\eta'-\epsilon'&\zeta_{22}^d&\eta+\epsilon\\0& \eta-\epsilon&1
\end{array}\right]{\cal M}_d^0\\
&\\
M_\nu^D=\left[
\begin{array}{ccc}
0&-3\epsilon'&0\\3\epsilon'&\zeta_{22}^u&\sigma-3\epsilon\\
0&\sigma+3\epsilon&1\end{array}\right]{\cal M}_u^0;& M_l=\left[
\begin{array}{ccc}
0&\eta'-3\epsilon'&0\\
\eta'+3\epsilon'&\zeta_{22}^d&\eta-3\epsilon\\0& \eta+3\epsilon&1
\end{array}\right]{\cal M}_d^0\\
\end{array}
\end{eqnarray}

These matrices are defined in the gauge basis and are multiplied
by $\bar\Psi_L$ on left and $\Psi_R$ on right. For instance, the
row and column indices of $M_u$ are given by $(\bar u_L, \bar c_L,
\bar t_L)$ and $(u_R, c_R, t_R)$ respectively. These matrices have
a hierarchical structure which can be attributed to a presumed
U(1)-flavor symmetry (see e.g. \cite{JCPKeK,BPR}), so that in
magnitudes 1 $\gg
\sigma\sim\eta\sim\epsilon\gg\zeta_{22}^u\sim\zeta_{22}^d\gg\eta'>\epsilon'$.
Following the constraints of SO(10) and the U(1)-flavor symmetry,
such a pattern of mass-matrices can be obtained using a {\it
minimal} Higgs system consisting of $\mathbf{45_{H}},
\mathbf{16_{H}}, \overline{\mathbf{16}}_{\mathbf{H}},
\mathbf{10_{H}}$ and a singlet {\it S} of SO(10)\footnote{Both the
BPW and the AB models bear similarities in the choice of the Higgs
system, yet there are significant differences in the mass
matrices. See text for details.}, which lead to effective
couplings of the form \cite{JCPKeK,BPR}:
\begin{eqnarray}
\label{eq:Yuk} {\cal L}_{\rm Yuk} &=& h_{33}{\bf 16}_3{\bf
16}_3{\bf 10}_H  + [ h_{23}{\bf 16}_2{\bf 16}_3{\bf
10}_H(S/M) \nonumber \\
&+& a_{23}{\bf 16}_2{\bf 16}_3{\bf 10}_H ({\bf
45}_H/M')(S/M)^p+g_{23}{\bf 16}_2{\bf 16}_3{\bf 16}_H^d ({\bf
16}_H/M'')(S/M)^q] \nonumber \\ &+& \left[h_{22}{\bf 16}_2{\bf
16}_2{\bf 10}_H(S/M)^2+g_{22}{\bf 16}_2{\bf 16}_2 {\bf
16}_H^d({\bf 16}_H/M'')(S/M)^{q+1} \right] \nonumber \\ &+&
\left[g_{12}{\bf 16}_1{\bf 16}_2 {\bf 16}_H^d({\bf
16}_H/M'')(S/M)^{q+2}+ a_{12}{\bf 16}_1{\bf 16}_2 {\bf 10}_H({\bf
45}_H/M')(S/M)^{p+2} \right]~.
\end{eqnarray}

The powers of ($S/M$) are determined by flavor-charge assignments
(see Refs. \cite{JCPKeK} and \cite{BPR}). The mass scales $M'$,
$M''$ and $M$ are of order $M_{\rm string}$ or (possibly) of order
$M_{GUT}$\cite{FN6}. Depending on whether $M'(M'')\sim M_{\rm
GUT}$ or $M_{\rm string}$ (see \cite{FN6}), the exponent $p(q)$ is
either one or zero \cite{FN8}. The VEVs of $\langle{\bf
45}_H\rangle$ (which is along $B-L$), $\langle{\bf
16}_H\rangle=\langle{\bf\overline {16}}_H\rangle$ (along
$\langle\tilde{\nu}_{RH}\rangle$) and $\langle S \rangle$ are of
the GUT-scale, while those of $\langle{\bf 10}_H\rangle$ and
$\langle{\bf 16}_H^d\rangle$ are of the electroweak scale
\cite{BPW,FN7}. The combination $ \mathbf{10_{H}}.\mathbf{45_{H}}$
effectively acts like a $\mathbf{120}$ which is antisymmetric in
family space and is along $B-L$. The hierarchical pattern is
determined by the suppression of the couplings by appropriate
powers of $M_{GUT}$/($M$, $M' or M''$). The entry ``1'' in the
matrices arises from the dominant $\mathbf{16_3 16_3 10_{\rm H}}$
term. The entries $\epsilon$ and $\epsilon'$ arising from the
$\mathbf{16_i 16_j 10_{\rm H}} 45_{\rm H}$ terms, are proportional
to $B-L$ and are antisymmetric in family space. Thus ($\epsilon,
\epsilon'$)$\to -3(\epsilon, \epsilon')$ as $q \to l$. The
parameter $\sigma$ comes from the $\mathbf{16_2 16_3 10_{\rm H}}$
term and contributes equally to the up and down sectors, whereas
$\hat\eta\equiv \eta-\sigma$, arising from $\mathbf{16_2 16_3
16_{\rm H}^{\rm d} 16_{\rm H}}$ operator, contributes only to the
down and charged lepton sectors. Similarly, $\zeta_{22}^u$ arises
from the $\mathbf{16_2 16_2 10_{\rm H}}$ term while $\zeta_{22}^d$
gets contributions from both $\mathbf{16_2 16_2 10_{\rm H}}$ and
$\mathbf{16_2 16_2 16_{\rm H}^{\rm d} 16_{\rm H}}$ operators.
Finally, $\eta'$, which is present only in the down and charged
lepton sectors, gets a contribution from $\mathbf{16_1 16_2
16_{\rm H}^{\rm d} 16_{\rm H}}$ terms in the Yukawa Lagrangian
(see Eq. (\ref{eq:Yuk})).

The right-handed neutrino masses arise from the effective
couplings of the form \cite{FN30}:
\begin{equation}
\label{eq:LMaj} \mathcal{L}_{\mathrm{Maj}} = f_{ij} \mathbf{16}_i
\mathbf{16}_j \overline{\mathbf{16}}_H \overline{\mathbf{16}}_H/M
\end{equation}

\noindent where the $f_{ij}$'s include appropriate powers of
$\langle S \rangle/M$. The hierarchical form of the Majorana
mass-matrix for the RH neutrinos is \cite{BPW}:
\begin{eqnarray}
\label{eq:MajMM} M_R^\nu=\left[
\begin{array}{ccc}
x & 0 & z \\
0 & 0 & y \\
z & y & 1
\end{array}
\right]M_R
\end{eqnarray}

Following flavor charge assignments (see \cite{JCPKeK}), we have
$1\gg y \gg z \gg x$. We expect $M_{st}\lsim M \lsim M_{Pl}$ where
$M_{st}\approx 4\times 10^{17}$ GeV and thus $M\approx 10^{18}$
GeV (1/2--2). The magnitude of M$_{\rm R}$ can now be estimated by
putting $f_{33}\approx 1,~\langle
\overline{\mathbf{16}}_H\rangle\approx 2\times 10^{16}$ GeV and
$M\approx (1/2-2)\ 10^{18}$ GeV \cite{BPW,JCPKeK}. This yields:
$M_R = f_{33} \langle \overline{\mathbf{16}}_H \rangle^2/M \approx
(4\times 10^{14}\mbox{ GeV})(1/2\mbox{--}2)$.

Thus the Majorana masses of the RH neutrinos are given
by\cite{JCPKeK,BPW}:
\begin{eqnarray}
\label{eq:MajM}
M_{3}& \approx & M_R\approx 4\times10^{14}\mbox{ GeV (1/2-2)},\nonumber\\
M_{2}& \approx & |y^2|M_{3}\approx \mbox{$10^{12}$ GeV(1/2-2)},\\
M_{1}& \approx & |x-z^2|M_{3} \sim (1/4\mbox{-}2)10^{-4}M_{3} \nonumber \\
 & & \sim 4\times 10^{10} \mbox{\ GeV}(1/8-4).\nonumber
\end{eqnarray}

\noindent Note that both the RH neutrinos as well as the light
neutrinos have hierarchical masses.

In the BPW model of Ref. \cite{BPW}, the parameters $\sigma, \eta,
\epsilon\ etc.$ were chosen to be real. Setting $\zeta_{22}^d =
\zeta_{22}^u = 0$, and with $m_t^{\rm phys}=174$ GeV,
$m_c(m_c)=1.37$ GeV, $m_s(1\mbox{ GeV})=110-116$ MeV, $m_u(1\mbox{
GeV})=6$ MeV, and the observed masses of $e$, $\mu$, and $\tau$ as
inputs, for this CP conserving case the following fit for the
parameters was obtained in Ref. \cite{BPW}:
\begin{eqnarray}
\label{eq:fit}
\begin{array}{l}
\ \sigma\approx 0.110, \quad \eta\approx 0.151, \quad
\epsilon\approx -0.095,
 \quad |\eta'|\approx 4.4 \times 10^{-3},\\
\begin{array}{l}
\epsilon'\approx 2\times 10^{-4},\quad {\cal M}_u^0\approx
m_t(M_X)\approx 100 \mbox{ GeV},\quad {\cal M}^0_d\approx
m_{\tau}(M_X)\approx 1.1 \mbox{ GeV}.
\end{array}
\end{array}
\end{eqnarray}
These output parameters remain stable to within 10\% corresponding
to small variations ($\lsim 10$\%) in the input parameters of
$m_{t}$, $m_{c}$, $m_{s}$, and $m_{u}$. These in turn lead to the
following predictions for the quarks and light neutrinos
\cite{BPW}, \cite{JCPKeK}:
\begin{eqnarray}
\label{eq:pred}
\begin{array}{l}
m_b(m_b) \approx (4.7\mbox{--}4.9) \mbox{\ GeV},\\
\sqrt{\Delta m_{23}^2} \approx m(\nu_3) \approx \mbox{(1/24 eV)(1/2--2)},\\
\begin{array}{lcl}
V_{cb} & \approx &
\left|\sqrt{\frac{m_s}{m_b}\left|\frac{\eta+\epsilon}
{\eta-\epsilon}\right|} - \sqrt{\frac{m_c}{m_t}\left|\frac{\sigma
+\epsilon}{\sigma-\epsilon}\right|}\right| \\
 & \approx & 0.044,
\end{array}\\
\left\{ \begin{array}{lcl}
\theta^{\mathrm{osc}}_{\nu_{\mu}\nu_{\tau}} & \approx &
\left|\sqrt{\frac{m_\mu}{m_\tau}} \left|
\frac{\eta-3\epsilon}{\eta+3\epsilon} \right|^{1/2} +
\sqrt{\frac{m_{\nu_2}}{m_{\nu_3}}}\right| \\
& \approx & |0.437+(0.378\pm 0.03)| {\mbox{ (for $\frac{m(\nu_2)}{m(\nu_3)}\approx 1/6$),}}\\
\multicolumn{3}{l}{\mbox{Thus, } \sin^2
2\theta^{\mathrm{osc}}_{\nu_{\mu}\nu_{\tau}}\approx 0.993,}
\end{array}\right.\\
V_{us}\approx
\left|\sqrt{\frac{m_d}{m_s}}-\sqrt{\frac{m_u}{m_c}}\right|
\approx 0.20,\\
\left|\frac{V_{ub}}{V_{cb}} \right|\approx
\sqrt{\frac{m_u}{m_c}}\approx
0.07,\\
m_d(\mbox{1 GeV})\approx \mbox{8 MeV}.
\end{array}
\end{eqnarray}

To allow for CP violation, this framework can be extended to
include phases for the parameters in Ref. \cite{BPR}. Remarkably
enough, it was found that there exists a class of fits within the
SO(10)/G(224) framework, which correctly describes not only  (a)
fermion masses, (b) CKM mixings and (c) neutrino oscillations
\cite{BPW,JCPKeK}, but also (d) the observed CP and flavor
violations in the K and B systems (see Ref. \cite{BPR} for the
predictions in this regard). A representative of this class of
fits (to be called fit A) is given by \cite{BPR}:
\begin{eqnarray}
\label{eq:fitA}
\begin{array}{l}
\ \ \sigma = 0.109-0.012i, \quad \eta = 0.122-0.0464i, \quad
\epsilon =
-0.103, \quad \eta' = 2.4 \times 10^{-3},\\
\begin{array}{l}
\ \epsilon' = 2.35\times 10^{-4}e^{i(69^{\circ})},\quad
\zeta_{22}^d = 9.8\times 10^{-3}e^{-i(149^{\circ})},\quad({\cal
M}_u^0,\ {\cal M}^0_d)\approx (100,\ 1.1) \mbox{ GeV}.
\end{array}
\end{array}
\end{eqnarray}

\noindent In this particular fit $\zeta_{22}^u$ is set to zero for
the sake of economy in parameters. However, allowing for
$\zeta_{22}^u\lsim (1/3)(\zeta_{22}^d)$ would still yield the
desired results. Because of the success of this class of fits in
describing correctly all four features (a), (b), (c) and (d)
mentioned above - which is a non-trivial feature by itself - we
will use fit A as a representative to obtain the sfermion
mass-insertion parameters $\hat{\delta}^{ij}_{LL,RR}$,
$(\delta^{ij}_{LL})^{RHN}$ and $\delta^{ij}_{LR,RL}$ in the lepton
sector and thereby the predictions of the BPW model and its
enxtension (Ref. \cite{BPR}) for lepton flavor violation.

The fermion mass matrices $M_u$, $M_d$ and $M_l$ are diagonalized
at the GUT scale $\approx 2\times 10^{16}$ GeV by bi-unitary
transformations:
 \begin{eqnarray}
 \label{eq:xdxu}
M_{u,d,l}^{diag}\ =\ X_L^{(u,d,l)\dagger}M_{u,d,l} X_R^{(u,d,l)}
\end{eqnarray}

\noindent The approximate analytic expressions for the matrices
$X^d_{L,R}$ can be found in \cite{BPR}. The corresponding
expressions for $X^l_{L,R}$ can be obtained by letting ($\epsilon,
\epsilon'$)$\to -3(\epsilon, \epsilon'$). For our calculations,
the mass-matrices have been diagonalized numerically.
\\
\\
{\bf The Albright-Barr Model}

The Dirac mass matrices of the u, d, l and $\nu$ sectors are given
by \cite{AB}:

\begin{eqnarray}
\label{eq:matAB}
\begin{array}{cc}
M_u=\left[
\begin{array}{ccc}
\tilde{\eta}&0&0\\0&0&\tilde{\epsilon}/3\\0&-\tilde{\epsilon}/3&1
\end{array}\right]{\cal M}_U;&
M_d=\left[
\begin{array}{ccc}
0&\tilde{\delta}&\tilde{\delta'}e^{i\phi}\\
\tilde{\delta}&0&\tilde{\sigma}+\tilde{\epsilon}/3\\\tilde{\delta'}e^{i\phi}&
-\tilde{\epsilon}/3&1
\end{array}\right]{\cal M}_D\\
&\\
M_\nu^D=\left[
\begin{array}{ccc}
\tilde{\eta}&0&0\\0&0&-\tilde{\epsilon}\\0&\tilde{\epsilon}&1\end{array}\right]{\cal
M}_U;& M_l=\left[
\begin{array}{ccc}
0&\tilde{\delta}&\tilde{\delta'}e^{i\phi}\\
\tilde{\delta}&0&-\tilde{\epsilon}\\\tilde{\delta'}e^{i\phi}&
\tilde{\sigma}+\tilde{\epsilon}&1
\end{array}\right]{\cal M}_D\\
\end{array}
\end{eqnarray}

These matrices are defined with the convention that the
left-handed fermions multiply them from the right, and the left
handed antifermions from the left. The AB model involves a
multitude of Higgs multiplets to generate fermion masses and
mixings including a $\mathbf{45_{\rm H}}$, two pairs of
$\mathbf{16_{\rm H} + \overline{16_{\rm H}}}$, two pairs of
$\mathbf{10_{\rm H}}$ and several singlets of SO(10). The ``1''
entry in the mass matrices arises from the dominant $\mathbf{16_3
16_3 10_H}$ operator. The $\tilde\epsilon$ entry arises from
operators of the form $\mathbf{16_2 16_3 10_H 45_H}$ (as in the
BPW model). Since $\langle\mathbf{45_H}\rangle \propto B-L$, the
$\tilde\epsilon$ entry is antisymmetric, and brings in a factor of
1/3 in the quark sector. The $\tilde\sigma$ term comes from the
operator $\mathbf{16_2 16_3 16_H 16'_H}$ by integrating out the
$\mathbf{10}$s of SO(10). (Note that the two $\mathbf{16}$s of
Higgs, $\mathbf{16_H}$ and $\mathbf{16'_H}$, are distinct). The
$\mathbf{16'_H}$ breaks the electroweak symmetry but does not
participate in the GUT scale breaking of SO(10). The resulting
operator is
$\mathbf{\overline{5}(16_2)10(16_3)\langle\overline{5}(16'_H)\rangle\langle
1(16_H)\rangle}$, where the $\mathbf{\overline{5}, 10}$ and
$\mathbf{1} \subset$ SU(5). Thus the $\tilde\sigma$ contributes
``lopsidedly'' to the l and d matrices. The entries $\tilde\delta$
and $\tilde\delta'$ arise from the operators $\mathbf{16_i 16_j
16_H 16'_H}$, like the $\tilde\sigma$ and contribute only to the l
and d matrices. Finally, $\tilde\eta$, which enters the u and
$\nu$ Dirac mass matrices, is of order $10^{-5}$ and arises from
higher dimensional operators. The Majorana mass matrix for the
right-handed neutrinos in the AB model is taken to have the
following form:

\begin{eqnarray}
\label{eq:MajMMAB} M_R=\left[
\begin{array}{ccc}
c^2 \tilde\eta^2 & -b \tilde\epsilon \tilde\eta & a \tilde\eta \\
-b \tilde\epsilon \tilde\eta & \tilde\epsilon ^2 & -\tilde\epsilon \\
a \tilde\eta & -\tilde\epsilon & 1
\end{array}
\right]\Lambda_R
\end{eqnarray}

\noindent with $\Lambda_R = 2.5\times 10^{14}$ GeV. The parameters
$a$, $b$ and $c$ are of order one to give the LMA solution for
neutrino oscillations. Given below is a fit to the parameters
$\tilde\sigma,\ \tilde\epsilon,\ \tilde\delta\ etc.$ which gives
the values of the fermion masses and the CKM elements in very good
agreement with observations \cite{ABPRD64,Jankowski}:
\begin{eqnarray}
\label{eq:fitAB}
\begin{array}{l}
\ \tilde\sigma = 1.78, \quad \tilde\epsilon = 0.145, \quad
\tilde\delta = 8.6\times 10^{-3}, \quad \tilde\delta' = 7.9 \times 10^{-3},\\
\begin{array}{l}
\phi = 126^{\circ}, \quad \tilde\eta=8\times 10^{-6}, \quad ({\cal
M}_u,\ {\cal M}_d)\approx (113,\ 1) \mbox{ GeV}.
\end{array}
\end{array}
\end{eqnarray}

In the next section, we turn to lepton flavor violation.

\section{The Three Sources of Lepton Flavor Violation}

As in Refs. \cite{BPR} and \cite{LFV}, we assume that
flavor-universal soft SUSY-breaking is transmitted to the
SM-sector at a messenger scale M$^*$, where M$_{GUT}< $ M$^{*}\le$
M$_{string}$. This may naturally be realized e.g. in models of
mSUGRA \cite{mSUGRA}, or gaugino-mediation \cite{gauginomed} or in
a class of anomalous U(1) D-term SUSY breaking models
\cite{anomU(1),faraggiJCP}. With the assumption of extreme
universality as in CMSSM, supersymmetry introduces five parameters
at the scale M$^{*}$:
\begin{center}
$m_o, m_{1/2}, A_o, \tan\beta\ {\rm and}\ sgn(\mu).$
\end{center}

\noindent For most purposes, we will adopt this restricted version
of SUSY breaking with the added restriction that $A_o$ = 0 at
M$^{*}$ \cite{gauginomed}. However, we will not insist on strict
Higgs-squark-slepton mass universality. Even though we have flavor
preservation at M$^{*}$, flavor violating scalar
(mass)$^2$--transitions arise in the model through RG running from
M$^*$ to the EW scale. As described below, we thereby have {\it
three sources} of lepton flavor violation \cite{BPR,LFV}.

\noindent {\bf (1) RG Running of Scalar Masses from M$^{*}$ to
M$_{\rm GUT}$.}

With family universality at the scale M$^{*}$, all sleptons have
the mass m$_o$ at this scale and the scalar (mass)$^2$ matrices
are diagonal. Due to flavor dependent Yukawa couplings, with $h_t
= h_b = h_\tau (= h_{33})$ being the largest, RG running from
M$^{*}$ to M$_{\rm GUT}$ renders the third family lighter than the
first two (see e.g. \cite{BHS}) by the amount:
\begin{eqnarray}
\label{eq:deltambr} \Delta\hat{m}_{\tilde{b}_{L}}^2 =
\Delta\hat{m}_{\tilde{b}_{R}}^2 =
\Delta\hat{m}_{\tilde{\tau}_{L}}^2 =
\Delta\hat{m}_{\tilde{\tau}_{R}}^2 \equiv\Delta\approx
\bigl(\frac{30m_o^2}{16\pi^2}\bigr) h_t^2\ ln(M^{*}/M_{GUT})~.
\end{eqnarray}
The factor 30$\to$12 for the case of G(224). The slepton
(mass)$^2$ matrix thus has the form $\tilde{\rm
M}^{(o)}_{\tilde{l}}$ = diag(m$_o^2$, m$_o^2$, m$_o^2 -\Delta$).
As mentioned earlier, the spin-1/2 lepton mass matrix is
diagonalized at the GUT scale by the matrices X$_{L,R}^l$.
Applying the same transformation to the slepton (mass)$^2$ matrix
(which is defined in the gauge basis), i.e. by evaluating
X$_L^{l\dagger}$($\tilde{\rm M}^{(o)}_{\tilde{l}}$)$_{LL}$ X$_L^l$
and similarly for L$\to$R, the transformed slepton (mass)$^2$
matrix is no longer diagonal. The presence of these off-diagonal
elements (at the GUT-scale) given by:
\begin{eqnarray}
\label{eq:deltahat} (\hat{\delta}_{LL,RR}^l)_{ij}=
\left(X_{L,R}^{l\dagger}(\tilde{\rm M}^{(o)}_{\tilde{l}})
X_{L,R}^l\right)_{ij}/m^2_{\tilde{l}}
\end{eqnarray}
\noindent induces flavor violating transitions
$\tilde{l}_{L,R}^i\to \tilde{l}_{L,R}^j$. Here m$_{\tilde{l}}$
denotes an average slepton mass and the hat signifies GUT-scale
values. Note that while the $(mass)^2$-shifts given in Eq.
(\ref{eq:deltambr}) are the same for the BPW and the AB models,
the mass insertions $\hat{\delta}_{LL,RR}$ would be different for
the two models since the matrices $X_{L,R}^l$ are different. As
mentioned earlier, the approximate analytic expressions for the
matrices $X^d_{L,R}$ for the BPW-model can be found in \cite{BPR}.
The corresponding expressions for $X^l_{L,R}$ can be obtained by
letting ($\epsilon, \epsilon'$)$\to -3(\epsilon, \epsilon'$),
though we use the exact numerical results in our calculations.

\noindent {\bf (2) RG Running of the $A-$parameters from M$^{*}$
to M$_{\rm GUT}$.}

Even if  $A_o$ = 0 at the scale M$^{*}$ (as we assume for
concreteness, see also \cite{gauginomed}). RG running from M$^{*}$
to M$_{\rm GUT}$ induces $A-$parameters at M$_{\rm GUT}$, invoving
the SO(10)/G(224) gauginos; these yield chirality flipping
transitions ($\tilde{l}^i_{L,R}\to \tilde{l}^j_{R,L}$). If we let
$M_{{\bf 16_H}}\approx M_{{\bf 10_H}}\approx M_{{\rm GUT}}$,
following the general analysis given in \cite{BHS}, the induced
$A-$parameter-matrix for the {\bf BPW} model is given by (see
\cite{LFV} for details):
\begin{eqnarray}
\label{eq:AdBPW} (A^l_{LR})_{\rm BPW} =Z\ ln(\frac{M^{*}}{M_{{\rm
GUT}}}) (X_L^l)^{\dagger}\left[
\begin{array}{ccc}
0&-285\epsilon'+90\eta' &0\\285\epsilon'+90\eta'
&90\zeta_{22}^d-27\zeta_{22}^u&-285\epsilon+90\eta-27\sigma\\0&
285\epsilon+90\eta-27\sigma&63
\end{array}\right] X_R^l
\end{eqnarray}

\noindent where $Z = \bigl(\frac{1}{16\pi^2}\bigr) h_t g_{10}^2
M_{\lambda}$. The coefficients
($\frac{63}{2},\frac{95}{2},\frac{90}{2}$) are the sums of the
Casimirs of the SO(10) representations of the chiral superfields
involved in the diagrams. For the case of G(224), we need to use
the substitutions:
($\frac{63}{2},\frac{95}{2},\frac{90}{2}$)$\to$($\frac{27}{2},\frac{43}{2},\frac{42}{2}$).
The X$^l_{L,R}$ are defined in Eq. (\ref{eq:deltahat}). The A-term
contribution is directly proportional to the SO(10) gaugino mass
$M_\lambda$ and thus to $m_{1/2}$. For approximate analytic
expressions of X$^l_{L,R}$, see Refs. \cite{BPR} and \cite{LFV}.

For the {\bf Albright-Barr} model, the induced $A-$matrix for the
leptons is given by:
\begin{eqnarray}
\label{eq:AdAB} (A^l_{LR})_{\rm AB} =Z\ ln(\frac{M^{*}}{M_{{\rm
GUT}}})(X_L^l)^{\dagger} \left[
\begin{array}{ccc}
0&90\tilde{\delta}&90\tilde{\delta'}e^{i\phi}\\
90\tilde{\delta}&0&-95\tilde{\epsilon}\\90\tilde{\delta'}e^{i\phi}&
90\tilde{\sigma}+95\tilde{\epsilon}&63
\end{array}\right] X_R^l
\end{eqnarray}
\noindent $(A^l_{LR})_{\rm AB}$ is transformed to the SUSY basis
by multiplying it with the matrices that diagonalize the lepton
mass matrix i.e. $X_{L,R}^l$ as in Eq. (\ref{eq:AdBPW}). The
chirality flipping transition angles are defined as :
\begin{eqnarray}
\label{eq:deltalr} (\delta^{l}_{LR})_{ij}\ \equiv\
(A^{l}_{LR})_{ij}\ \bigl(\frac{v_d}{m_{\tilde{l}}^2}\bigr)\ =\
(A^{l}_{LR})_{ij}\ \bigl(\frac{v_u}{\tan\beta\
m_{\tilde{l}}^2}\bigr)~.
\end{eqnarray}

\noindent {\bf (3) RG Running of scalar masses from M$_{\rm GUT}$
to the RH neutrino mass scales:}

We work in a basis in which the charged lepton Yukawa matrix Y$_l$
and M$_R^{\nu}$ are diagonal at the GUT scale. The off-diagonal
elements in the Dirac neutrino mass matrix Y$_N$ in this basis
give rise to lepton flavor violating off-diagonal components in
the left handed slepton mass matrix through the RG running of the
scalar masses from M$_{\rm GUT}$ to the RH neutrino mass scales
M$_{R_i}$ \cite{60BM}. The RH neutrinos decouple below M$_{R_i}$.
(For RGEs for MSSM with RH neutrinos see e.g. Ref. \cite{HNTY}).
In the leading log approximation, the off-diagonal elements in the
left-handed slepton (mass)$^2$-matrix, thus arising, are given by:
\begin{eqnarray}
\label{eq:RHN} (\delta_{LL}^l)_{ij}^{RHN} =
\frac{-(3m_o^2+A_o^2)}{8\pi^2}\sum_{k=1}^{3}
(Y_N)_{ik}(Y_N^{*})_{jk}\ ln(\frac{M_{GUT}}{M_{R_k}})~.
\end{eqnarray}

\noindent The superscript RHN denotes the contribution due to the
presence of the RH neutrinos. For the case of the AB-model, in the
above expression, $(Y_N)_{ik}(Y_N^{*})_{jk} \to
(Y_N)_{kj}(Y_N^{*})_{ki}$ because of the definition of the
mass-matrices. The masses M$_{R_i}$ of RH neutrinos are determined
from Eqs. (\ref{eq:MajM}) and (\ref{eq:MajMMAB}) for the BPW and
AB models respectively. The total LL contribution, including
post-GUT contribution (Eq. (\ref{eq:deltahat})) and the RH
neutrino contribution (Eq. (\ref{eq:RHN})), is thus:
\begin{eqnarray}
\label{eq:deltatot}
(\delta_{LL}^l)_{ij}^{Tot}=(\hat{\delta}_{LL}^l)_{ij}+(\delta_{LL}^l)_{ij}^{RHN}
\end{eqnarray}

We will see in the next section that this contribution to $\mu\to
e\gamma$ is very different in the two models (noted in part in
Ref. \cite{Barr}) and provides a way to distinguish the two
models. We find that this contribution in the AB model is a factor
of $\sim 25-35$ larger in the the amplitude than that in the BPW
model, and this difference arises entirely due to the structure of
the mass matrices. We also find that this difference in the mass
matrices, also gives rise to large differences in the edm of the
electron between the two models.

We now present some results on lepton flavor violation. In the
following section we will turn to CP violation, and see how the
two models compare.

\section{Results on Lepton Flavor Violation}
The decay rates for the lepton flavor violating processes l$_i\to$
l$_j \gamma\ (i>j)$ are given by:
\begin{eqnarray}
\label{eq:br} \Gamma(l_i^+ \to l_j^+ \gamma) =\frac{e^2
m_{l_i}^3}{16\pi} \left(|A_L^{ji}|^2 + |A_R^{ji}|^2\right)
\end{eqnarray}

\noindent Here $A_L^{ji}$ is the amplitude for $(l_i)^+_L
\rightarrow (l_j)^+ \gamma$ decay, while $A_R^{ji} = Amp((l_i)^+_R
\rightarrow (l_j)^+ \gamma)$. The amplitudes $A_{L,R}^{ji}$ are
evaluated in the mass insertion approximation using the
$(\delta_{LL}^l)^{Tot},\ \delta_{RR}^l$ and $\delta_{LR,RL}^l$
calculated as above. The general expressions for the amplitudes
$A_{L,R}^{ji}$ in one loop can be found in e.g. Refs. \cite{HNTY}
and \cite{HN}. We include the contributions from both chargino and
neutralino loops with or without the $\mu-$term.

In Table 1 we give the branching ratio of the process $\mu\to
e\gamma$ and the individual contributions from the sources
$\hat\delta^{ji}_{LL},\ \delta^{ji}_{LR,RL}$ and
$(\delta^{ji}_{LL})^{RHN}$ (see Eqs. (\ref{eq:deltahat}),
(\ref{eq:deltalr}) and (\ref{eq:RHN})) evaluated in the SO(10)-BPW
model, with some sample choices of (m$_o$, m$_{1/2}$). For these
calculations, to be concrete, we set
ln$\left(\frac{M^{*}}{M_{GUT}}\right)$ = 1, i.e. $M^{*}\approx 3
M_{GUT}$, $\tan\beta = 10$, $A_o$( at M$^*$) = 0 and $\mu>0$. In
the BPW model, for concreteness, the RH neutrino masses are taken
to be M$_{R_1} = 10^{10}$ GeV, M$_{R_2} = 10^{12}$ GeV and
M$_{R_3} = 5\times 10^{14}$ GeV (see Eq. (\ref{eq:MajM})). For the
masses of the right-handed neutrinos in the AB model, we set
M$_{R_1} = 7.5\times 10^{8}$ GeV, M$_{R_2} = 7.5 \times 10^{8}$
GeV and M$_{R_3} = 2.6\times 10^{14}$ GeV corresponding to $a = c
= 4$ and $b = 6$ in Eq. (\ref{eq:MajMMAB}). (The results on the
rate of $\mu\to e\gamma$, presented in the following table do not
change very much for other ($\mathcal{O}(1)$) values of $ a,\ b$
and $c$.). It should be noted that {\it the corresponding values
for the G(224)-BPW model are smaller than those for the SO(10)-BPW
model approximately by a factor of 4 to 6 in the rate, provided
$\ln(M^*/M_{GUT})$ is the same in both cases (see comments below
Eqs. (\ref{eq:deltambr}) and (\ref{eq:AdBPW}))}. A pictorial
representation of these results is depicted in Figs. 1 and 2.

\noindent
\begin{tabular}{|c|c|c|c|c|c|}
\hline
\rule[-3mm]{0mm}{8mm}$(m_o,~m_{1/2})$(GeV)&$A_L^{(1)}(\hat{\delta}_{LL})
$&
  $A_L^{(2)}(\delta_{LR})$&
  $A_R(\delta_{RL})$&$A_L^{(3)}((\delta_{LL})^{RHN})$&Br($\mu\to e\gamma$)\\
 \hline
 \hline
 (100, 250) BPW&$-1.2\times 10^{-10}$&$4.5\times
 10^{-13}$&$-7.2\times 10^{-11}$&$3.7\times 10^{-14}$&$1.3\times
 10^{-7}$\\
 \hline
 (100, 250) AB&$-8.5\times 10^{-11}$&$1.9\times
 10^{-12}$&$-6.4\times 10^{-11}$&$1.3\times 10^{-12}$&$8.0\times
 10^{-8}$\\
 \hline
 \hline
 (500, 250) BPW&$-1.9\times 10^{-12}$&$1.0\times
 10^{-12}$&$-1.6\times 10^{-12}$&$8.5\times 10^{-14}$&$2.2\times
 10^{-11}$\\
 \hline
 (500, 250) AB&$-1.4\times 10^{-12}$&$4.4\times
 10^{-12}$&$-1.4\times 10^{-12}$&$2.9\times 10^{-12}$&$2.6\times
 10^{-10}$\\
 \hline
 \hline
 {\bf (800, 250) BPW}&$-3.5\times 10^{-13}$&$6.1\times
 10^{-13}$&$-2.9\times 10^{-13}$&$4.9\times 10^{-14}$&$\mathbf{1.3\times
 10^{-12}}$\\
 \hline
 (800, 250) AB&$-2.6\times 10^{-13}$&$2.5\times
 10^{-12}$&$-2.6\times 10^{-13}$&$1.7\times 10^{-12}$&$1.1\times
 10^{-10}$\\
 \hline
 \hline
 {\bf (1000, 250) BPW}&$-1.5\times 10^{-13}$&$4.3\times
 10^{-13}$&$-1.2\times 10^{-13}$&$3.5\times 10^{-14}$&$\mathbf{8.1\times
 10^{-13}}$\\
 \hline
(1000, 250) AB&$-1.1\times 10^{-13}$&$1.8\times
 10^{-12}$&$-1.1\times 10^{-13}$&$1.2\times 10^{-12}$&$5.9\times
 10^{-11}$\\
 \hline
 \hline
 {\bf (600, 300) BPW}&$-1.3\times 10^{-12}$&$7.2\times
 10^{-13}$&$-1.1\times 10^{-12}$&$5.9\times 10^{-14}$&$\mathbf{1.1\times
 10^{-11}}$\\
 \hline
(600, 300) AB&$-9.8\times 10^{-13}$&$3.0\times
 10^{-12}$&$-9.7\times 10^{-13}$&$2.0\times 10^{-12}$&$1.3\times
 10^{-10}$\\
 \hline
 \hline
 (100, 500) BPW&$-5.4\times 10^{-11}$&$3.5\times
 10^{-14}$&$-2.8\times 10^{-11}$&$2.8\times 10^{-15}$&$2.6\times
 10^{-8}$\\
 \hline
(100, 500) AB&$-4.0\times 10^{-11}$&$1.5\times
 10^{-13}$&$-2.5\times 10^{-11}$&$9.7\times 10^{-14}$&$1.6\times
 10^{-8}$\\
 \hline
\hline
 (500, 500) BPW&$-4.3\times 10^{-12}$&$3.1\times
 10^{-13}$&$-3.3\times 10^{-12}$&$2.5\times 10^{-14}$&$1.9\times
 10^{-10}$\\
 \hline
(500, 500) AB&$-3.2\times 10^{-12}$&$1.3\times
 10^{-12}$&$-3.0\times 10^{-12}$&$8.6\times 10^{-13}$&$7.5\times
 10^{-11}$\\
 \hline
 \hline
 {\bf (1000, 500) BPW}&$-4.8\times 10^{-13}$&$2.6\times
 10^{-13}$&$-3.9\times 10^{-13}$&$2.1\times 10^{-14}$&$\mathbf{1.4\times
 10^{-12}}$\\
 \hline
(1000, 500) AB&$-3.5\times 10^{-13}$&$1.1\times
 10^{-12}$&$-3.5\times 10^{-13}$&$7.3\times 10^{-13}$&$1.6\times
 10^{-11}$\\
 \hline
 \hline

 (200, 1000) BPW&$-1.3\times 10^{-11}$&$8.8\times
 10^{-15}$&$-7.1\times 10^{-12}$&$7.2\times 10^{-16}$&$1.6\times
 10^{-9}$\\
 \hline
(200, 1000) AB&$-9.9\times 10^{-12}$&$3.7\times
 10^{-14}$&$-6.4\times 10^{-12}$&$2.4\times 10^{-14}$&$1.0\times
 10^{-9}$\\
 \hline
 \hline
 {\bf (1000, 1000) BPW}&$-1.1\times 10^{-12}$&$7.7\times
 10^{-14}$&$-8.3\times 10^{-13}$&$6.3\times 10^{-15}$&$\mathbf{1.2\times
 10^{-11}}$\\
 \hline
{\bf (1000, 1000) AB}&$-7.9\times 10^{-13}$&$3.2\times
 10^{-13}$&$-7.4\times 10^{-13}$&$2.2\times 10^{-13}$&$\mathbf{4.7\times
 10^{-12}}$\\
 \hline
\end{tabular}
\vskip.10in \noindent Table 1. {\small Comparison between the AB
and the BPW models of the various contributions to the amplitude
and of the branching ratio for $\mu\to e\gamma$ for the case of
SO(10). Each of the entries for the amplitudes should be
multiplied by a common factor a$_o$. Imaginary parts being small
are not shown. Only the cases shown in {\bf bold} typeface are in
accord with experimental bounds; the other ones are excluded. The
first three columns denote contributions to the amplitude from
post-GUT physics arising from the regime of $M^*\to M_{GUT}$ (see
Eqs. (\ref{eq:deltahat})--(\ref{eq:deltalr})), where for
concreteness we have chosen $\ln(M^*/M_{GUT})=1$. The fifth column
denotes the contribution from the right-handed neutrinos (RHN).
Note that the entries corresponding to the RHN-contribution are
much larger in the AB-model than those in the BPW-model; this is
precisely because the AB-model is lopsided while the BPW model is
hierarchical (see text). Note that for the BPW model, the post-GUT
contribution far dominates over the RHN-contribution while for the
AB model they are comparable. The last column gives the branching
ratio of $\mu\to e\gamma$ including contributions from all four
columns. The net result is that the AB  model is compatible with
the empirical limit on $\mu\to e\gamma$ only for rather heavy SUSY
spectrum like $(m_o,~ m_{1/2})\gsim (1000,~1000)$ GeV, whereas the
BPW is fully compatible with lighter SUSY spectrum like $(m_o,~
m_{1/2})\sim (600,~300)$ GeV (see text) for the case of SO(10),
and $(m_o,~ m_{1/2})\sim (400,~250)$ GeV for G(224). These results
are depicted graphically in Figs. 1 and 2. } \vspace*{5pt}

Before discussing the features of this table, it is worth noting
some distinguishing features of the BPW and the AB models. As can
be inferred from Eqs. (\ref{eq:AdBPW}) and (\ref{eq:AdAB}), for a
given $m_o$, the post-GUT contribution for both the BPW and the AB
models increases with increasing $m_{1/2}$ primarily due to the
A-term contribution. It turns out that for $m_{1/2}\gsim 300$ GeV,
this contribution becomes so large that Br($\mu\to e\gamma$)
exceeds the experimental limit, unless one chooses $m_o\gsim 1000$
GeV, so that the rate is suppressed due to large slepton masses.
This effect applies to both models.

For the hierarchical BPW model, however, it turns out that the RHN
contribution is strongly suppressed both relative to that in the
lopsided AB-model; and also relative to the post-GUT contributions
(see discussion below). As a result the dominant contribution for
the BPW model comes only from post-GUT physics, which decreases
with decreasing $m_{1/2}$ for a fixed $m_o$. Such a dependence on
$m_{1/2}$ is not so striking, however, for the AB model because in
this case, owing to the lopsided structure, the RHN contribution
(which is not so sensitive to $m_{1/2}$) is rather important and
is comparable to the post-GUT contribution.

Tables 1 and 2 bring out some very interesting distinctions
between the two models:

(1) The experimental limit on $\mu\to e\gamma$ is given by:
Br($\mu\to e\gamma) < 1.2\times 10^{-11}$ \cite{mue}. This means
that for the case of the AB model, with dominant contribution
coming not only from post-GUT physics but also from the RHN
contribution, only rather heavy SUSY spectrum, $(m_o,\
m_{1/2})\gsim (1000,\ 1000)$ GeV, is allowed. The BPW-model, on
the other hand, allows for relatively low $m_{1/2}$ ($\lsim 300$
GeV), with moderate to heavy $m_o$, which can be as low as about
600 GeV with $m_{1/2}\le 300$ GeV. {\it As a result, whereas the
AB model is consistent with $\mu\to e\gamma$ only for rather heavy
sleptons ($\gsim 1200$ GeV) and heavy squarks ($\gsim$ 2.8 TeV),
the BPW model is fully compatible with much lighter slepton masses
$\sim 600$ GeV, with squarks being 800 GeV to 1 TeV.} These
results hold for the case od SO(10). For the G(224) case the BPW
model would be consistent with the experimental limit on the rate
of $\mu\to e\gamma$ for even lighter SUSY spectrum including
values of $(m_o,~m_{1/2})\approx (400,~250)$ GeV, which
corresponds to $m_{\tilde{q}}\sim 780$ GeV and $m_{\tilde{l}}\sim
440$ GeV.

(2) From the point of view of forthcoming experiments we also note
that $\mu\to e\gamma$ for the BPW case, ought to seen with an
improvement in the current limit by a factor of 10--50. For the AB
case, even with a rather heavy SUSY spectrum  ($(m_o,\
m_{1/2})\gsim (1000,\ 1000)$ GeV), $\mu\to e\gamma$ should be seen
with an improvement by a factor of only 3--5. Such experiments are
being planned at the MEG experiment at PSI \cite{mueupcoming}

(3) As has been noted earlier in \cite{Barr} and more recently in
\cite{LFV}, the contribution to $A_L (\mu\to e\gamma)$ due to RH
neutrinos in the BPW model is approximately proportional to
$\eta-\sigma\approx 0.041$, which is naturally small because the
entries $\eta$ and $\sigma$ are of $\mathcal{O}(1/10)$ in
magnitude due to the hierarchical structure. In the AB-model on
the other hand, this contribution is proportional to
$\tilde{\sigma}+2\tilde\epsilon/3\approx 1.8$. Thus we expect that
in amplitude, the RHN contribution in the BPW model is smaller by
about a factor of 40 than that in the AB model. This has two
consequences:

(a) First, there is a dramatic difference between the two models
which becomes especially prominent if one drops the post-GUT
contribution, that amounts to setting $M^*= M_{GUT}$. In this case
the contribution to $(\mu\to e\gamma)$ comes entirely from the RHN
contribution. In this case  the branching ratio of $(\mu\to
e\gamma)$ in the two models differs by a factor of about
$(40)^2\sim \mathcal{O}(10^3)$ as depicted in table 2.

\vspace*{12pt}

\begin{tabular}{|c|c|c|}
\hline \rule[-3mm]{0mm}{8mm}$(m_o,~m_{1/2})$(GeV)&Br$(\mu\to
e\gamma)_{AB}^{RHN}$&Br$(\mu\to e\gamma)_{BPW}^{RHN}$\\ \hline
 (100, 250)&$1.2\times 10^{-11}$&$9.7\times
 10^{-15}$\\
 \hline
 (800, 250)&$2.1\times 10^{-11}$&$1.7\times
 10^{-14}$\\
 \hline
 (600, 300)&$2.8\times 10^{-11}$&$2.5\times
 10^{-14}$\\
 \hline
 (500, 500)&$5.3\times 10^{-12}$&$4.4\times
 10^{-15}$\\
 \hline
 (1000, 1000)&$3.4\times 10^{-13}$&$2.8\times
 10^{-16}$\\
 \hline
\end{tabular}
\vskip.10in \noindent Table 2. {\small Branching ratio for
$(\mu\to e\gamma)$ based only on the RHN contribution (this
corresponds to setting $M^*= M_{GUT}$) for the AB and BPW models
for different choices of $(m_o,~m_{1/2})$.} \vspace*{5pt}

\noindent It can be seen from table 2 that with only the RHN
contribution (which would be the total contribution if $M^*=
M_{GUT}$), the AB model is consistent with the limit on $\mu\to
e\gamma$ for light SUSY spectrum, e.g. for $(m_o,~m_{1/2})=(100,
250)$ GeV. A similar analysis for the AB model was done in Ref.
\cite{Jankowski} (including the RHN contribution only), and our
results agree with those of Ref. \cite{Jankowski}. One may expect
that for the same value of $m_{1/2}$, increasing $m_o$ would
result in decreasing the branching ratio. For example, from Eq.
(\ref{eq:RHN}), one may expect the rate for $\mu\to e\gamma$ to be
proportional to $(m_o^2/m_{\tilde{l}}^4)^2\sim 1/m_o^4$. However,
the associated loop function (see e.g. Ref. \cite{HN}) alters the
dependence on $(m_o,~m_{1/2})$ drastically; it increases with
increasing $m_o$, for fixed $m_{1/2}$. {\it The net result of
these two effects is that for the same $m_{1/2}$, a low $m_o\sim
100$ GeV and a high $m_o\sim 1000$ GeV, give nearly the same value
of the branching ratio for $\mu\to e\gamma$ with the inclusion of
only the RH neutrino contribution} (see Fig. 3) . This can also be
seen in the results of Ref. \cite{Jankowski} which analyzes the AB
model. The RHN contribution in the case of the BPW model is
extremely small because of its hierarchical structure, as
explained above.

Of course, in the context of supersymmetry breaking as in mSUGRA
or gaugino-mediation, we expect $M^*>M_{GUT}$, thus post-GUT
contributions should be included at least in these cases. With the
inclusion of post-GUT physics,as mentioned above, the AB model is
consistent with the experimental limit on $\mu\to e\gamma$, only
for very heavy SUSY spectrum with $(m_o,\ m_{1/2})\gsim (1000,\
1000)$ GeV, i.e. $m_{\tilde{l}}\gsim 1200$ GeV and
$m_{\tilde{q}}\gsim 2.8$ TeV; whereas the BPW model is fully
compatible with the empirical limit for significantly lower values
of $(m_o,\ m_{1/2})\sim (600,\ 300)$ GeV, i.e. $m_{\tilde{l}}\sim
600$ GeV and $m_{\tilde{q}}\sim 1$ TeV (see table 1).

(b) Second, it was shown in Ref. \cite{LFV} that the P-odd
asymmetry parameter for the process $(\mu^+\to e^+\gamma)$ defined
as $\mathcal{A}(\mu^+ \rightarrow e^+ \gamma) = (|A_L|^2 - |A_R|^2
)/ (|A_L|^2 + |A_R|^2)$ (where
$|A_L|=|A_L^{(1)}(\hat{\delta}_{LL})+A_L^{(2)}(\delta_{LR})+A_L^{(3)}|)$,
is typically negative for the BPW model except for cases with very
large $m_{1/2}$ e.g. $(m_o,\ m_{1/2})= (1000,\ 1000)$ or $(500,\
500)$ GeV. For the AB-case, due to the large RHN contribution,
$|A_L|>|A_R|$ and therefore the P-odd asymmetry parameter
${\mathcal A}$ would typically be positive. Thus the determination
of ${\mathcal A}$ in future experiments can help distinguish
between the BPW and the AB models.

For the sake of completeness, we give the branching ratios of the
processes $\tau\to\mu\gamma$ and $\tau\to e\gamma$ calculated in
the two models in table 3.

\vspace*{12pt}

\begin{tabular}{|c|c|c|c|c|}
\hline \rule[-3mm]{0mm}{8mm} ($m_o,\ m_{1/2}$)(GeV) &
\multicolumn{2}{c|}{AB-model}&\multicolumn{2}{c|} {BPW-model}\\
\hline
& Br($\tau\to\mu\gamma$)&  Br($\tau\to e\gamma$) &  Br($\tau\to\mu\gamma$) &  Br($\tau\to e\gamma$) \\

\hline

(100, 250)& $2.9\times 10^{-9}$ & $3.8\times 10^{-11}$
         & 2.6$\times 10^{-7}$  & 1.6$\times10^{-9}$
                  \\ \hline
(800, 250)& $1.0\times 10^{-8}$ & $4.5\times 10^{-11}$
         & 1.6$\times 10^{-9}$  & 6.8$\times10^{-12}$
                  \\ \hline
(600, 300)& $1.4\times 10^{-8}$ & $6.4\times 10^{-11}$
         & 2.1$\times 10^{-9}$  & 8.4$\times10^{-12}$
                  \\ \hline
(500, 500)& $2.4\times 10^{-9}$ & $1.0\times 10^{-11}$
         & 3.9$\times 10^{-10}$  & 1.8$\times10^{-12}$
                  \\ \hline
(1000, 1000)& $1.5\times 10^{-10}$ & $6.5\times 10^{-13}$
         & 2.5$\times 10^{-11}$  & 1.1$\times10^{-13}$
                  \\ \hline
\end{tabular}
\vskip.10in \noindent Table 3. {\small Branching ratios for
$(\tau\to \mu\gamma)$ and $(\tau\to e\gamma)$
 evaluated in the two models for the case of SO(10), for some sample
choices of $(m_o,~m_{1/2})$. We have set $\tan\beta = 10,\ \mu>0$
and ln$\left(\frac{M^{*}}{M_{GUT}}\right)$ = 1.} \vspace*{5pt}

From table 3 we see that the predictions for the branching ratios
for $(\tau\to \mu\gamma)$ and $(\tau\to e\gamma)$ in either model
are well below the current experimental limits. The process
$(\tau\to \mu\gamma)$ can be probed at BABAR and BELLE or at LHC
in the forthcoming experiments; $(\tau\to e\gamma)$ seems to be
out of the reach of the upcoming experiments.

In the following section we turn to CP violation in the two
models.

\section{Results on Fermion Masses, CKM Elements and CP Violation}
CP violation in the BPW model was studied in detail in Ref.
\cite{BPR}. We will recapitulate some of those results and do a
comparative study with the AB model. For any choice of the
parameters in the mass matrices ($\eta,\ \sigma,\ \epsilon$ {\it
etc.} for the BPW case, and $\tilde\sigma,\ \tilde\epsilon$ {\it
etc.} for the AB case), one gets the SO(10)-model based values of
$\rho_W$ and $\eta_W$, which generically can differ widely from
the SM-based phenomenological values. We denote the former by
$(\rho'_W)_{BPW,AB}$ and $(\eta'_W)_{BPW,AB}$ and the
corresponding contributions from the SM-interactions (based on
$\rho'_W$ and $\eta'_W$) by SM$'$. In our calculations we include
both the SM$'$ contribution and the SUSY contributions involving
the sfermion $(mass)^2$-parameters($\delta ^{ij}_{LL,RR,LR})$
which are in general CP violating. These parameters are completely
determined in each of the two models for a given choice of flavor
preserving SUSY-parameters (i.e. $m_o,\ m_{1/2},\ \mu$, and
$\tan\beta$; we set $A_o$ = 0 at $M^{*}$). Using the fits given in
Eqs. (\ref{eq:fitA}) and (\ref{eq:fitAB}), we get the following
values for the CKM elements and fermion masses using
$m_t(m_t)=167\ GeV$ and $m_{\tau}(m_{\tau})=1.777\ GeV$ as inputs:
\\ {\bf BPW:}
\begin{eqnarray}
\label{eq:predCKM}
\begin{array}{l}
\ \ \ \ ((V_{us},\ V_{cb},\ |V_{ub}|,\ |V_{td}|)(\le
m_Z))_{BPW}\approx
(0.2250,\ 0.0412,\ 0.0037,\ 0.0086)\\
\begin{array}{l}
\ \ \ (\bar\rho'_W)_{BPW} = 0.150,\quad (\bar\eta'_W)_{BPW} = 0.374\\
\begin{array}{l}
\ \ (m_b(m_b),\ m_c(m_c))\approx (4.97,\ 1.32)\
GeV\\
\begin{array}{l}
\ (m_s(1GeV),\ m_{\mu})\approx (101,\ 109)\
MeV\\
\begin{array}{l}
(m_u^{\circ}(1GeV),\ m_d^{\circ}(1GeV),\ m_e^{\circ})\approx
(10.1,\ 3.7,\ 0.13)\ MeV
\end{array}
\end{array}
\end{array}
\end{array}
\end{array}
\end{eqnarray}

\noindent {\bf AB:}
\begin{eqnarray}
\label{eq:predCKM_AB}
\begin{array}{l}
\ \ \ \ ((V_{us},\ V_{cb},\ |V_{ub}|,\ |V_{td}|)(\le
m_Z))_{AB}\approx
(0.220,\ 0.041,\ 0.0032,\ 0.0081)\\
\begin{array}{l}
\ \ \ (\bar\rho'_W)_{AB} = 0.148,\quad (\bar\eta'_W)_{AB} = 0.309\\
\begin{array}{l}
\ \ (m_b(m_b),\ m_c(m_c))\approx (4.97,\ 1.15)\
GeV\\
\begin{array}{l}
\ (m_s(1GeV),\ m_{\mu})\approx (177,\ 106)\
MeV\\
\begin{array}{l}
(m_u^{\circ}(1GeV),\ m_d^{\circ}(1GeV),\ m_e^{\circ})\approx
(3.2,\ 8.5,\ 0.56)\ MeV
\end{array}
\end{array}
\end{array}
\end{array}
\end{array}
\end{eqnarray}

The predictions of both models for the CKM elements are in good
agreement with the measured values, and $(\bar\rho'_W)$ and
$\bar\eta'_W)$ are close to the SM values in each case. It was
remarked in Ref. \cite{BPR} that for the BPW model, the masses of
the light fermions (u, d and e) can be corrected by allowing for
$\mathcal{O}(10^{-4}-10^{-5})$ ``11'' entries in the mass matrices
which can arise naturally through higher dimensional operators.
Such small entries will not alter the predictions for the CKM
mixings. For the AB model, the masses of the bottom and strange
quarks have been lowered by the gluino loop contributions from
5.12 GeV and 183 MeV to 4.97 GeV and 177 MeV respectively. Thus
from Eqs. (\ref{eq:predCKM}) and (\ref{eq:predCKM_AB}), we see
that both models are capable of yielding the gross pattern of
fermion masses and especially the CKM mixings in good accord with
observations; at the same time $(\bar\rho'_W)$ and $\bar\eta'_W)$
are close to the phenomenological SM values.

We now present some results on CP violation. We include both the
SM$'$ and the SUSY contributions in obtaining the total
contributions (denoted by ``Tot''). The SUSY contribution is
calculated using the squark mixing elements,
$\delta^{ij}_{LL,RR,LR}$, which are completely determined in both
models for any given choice of the SUSY breaking parameters $m_o,\
m_{1/2},\ A_o, \tan\beta$ and $sgn(\mu)$. As emphasized earlier,
in our calculations, the $\delta^{ij}s$ include contributions from
both post-GUT physics as well as those coming from RG running in
MSSM below the GUT scale. (For details, see Ref. \cite{BPR}). We
set $A_o = 0$ for concreteness, as before. Listed below in Table 4
are the results on CP and flavor violations in the
$\rm{K}^\circ-\overline{\rm{K}^\circ}$ and
$\rm{B_d}^\circ-\overline{\rm{B_d}^\circ}$ systems for the two
models. For these calculations we set $\ln(M^*/M_{GUT})=1$.

\noindent \vspace*{3pt}
\begin{tabular}{|c|c|c|c|c|c|}
\hline \rule[-3mm]{0mm}{8mm}$(m_o,~m_{1/2})$(GeV)&$\Delta
m_K^{s.d.}$(GeV)&
 $\epsilon_K (SM')$&
  $\epsilon_K (Tot)$&$\Delta m_{B_d}$ (GeV)&$S_{\psi K_S}$\\
 & Tot $\approx$ SM$'$ & & & Tot $\approx$ SM$'$& Tot $\approx$
 SM$'$\\
 \hline
 \hline
 (300, 300) BPW&$2.9\times 10^{-15}$&$2.8\times
 10^{-3}$&$2.6\times 10^{-3}$&$3.5\times 10^{-13}$&$0.73$\\
 \hline
 (300, 300) AB&$2.8\times 10^{-15}$&$2.2\times
 10^{-3}$&$2.2\times 10^{-3}$&$3.1\times 10^{-13}$&$0.66$\\
 \hline
 \hline
 (600, 300) BPW&$2.9\times 10^{-15}$&$2.8\times
 10^{-3}$&$2.0\times 10^{-3}$&$3.6\times 10^{-13}$&$0.73$\\
 \hline
 (600, 300) AB&$2.8\times 10^{-15}$&$2.2\times
 10^{-3}$&$1.4\times 10^{-3}$&$3.1\times 10^{-13}$&$0.66$\\
 \hline
 \hline
 (1000, 250) BPW&$2.9\times 10^{-15}$&$2.8\times
 10^{-3}$&$1.4\times 10^{-3}$&$3.6\times 10^{-13}$&$0.74$\\
 \hline
 (1000, 250) AB&$2.8\times 10^{-15}$&$2.2\times
 10^{-3}$&$-4.0\times 10^{-3}$&$3.13\times 10^{-13}$&$0.656$\\
 \hline
 \hline
 (1000, 500) BPW&$2.9\times 10^{-15}$&$2.83\times
 10^{-3}$&$2.6\times 10^{-3}$&$3.6\times 10^{-13}$&$0.73$\\
 \hline
 (1000, 500) AB&$2.8\times 10^{-15}$&$2.2\times
 10^{-3}$&$2.0\times 10^{-3}$&$3.1\times 10^{-13}$&$0.66$\\
 \hline
 \hline
  (1000, 1000) BPW&$2.9\times 10^{-15}$&$2.8\times
 10^{-3}$&$2.9\times 10^{-3}$&$3.5\times 10^{-13}$&$0.72$\\
 \hline
 (1000, 1000) AB&$2.8\times 10^{-15}$&$2.2\times
 10^{-3}$&$2.3\times 10^{-3}$&$3.1\times 10^{-13}$&$0.66$\\
 \hline
\end{tabular}
\vskip.10in \noindent Table 4. {\small CP violation in the
$K^\circ-\overline{K^\circ}$ and $B_d-\overline{B_d}$ systems as
predicted in the BPW and the AB models for some sample choices of
$(m_o,\ m_{1/2})$ and a generic fit of parameters (see
Eq.(\ref{eq:fitA}) for the BPW case and Eq. (\ref{eq:fitAB}) for
the AB case). The superscript s.d. on $\Delta m_K$ denotes the
short distance contribution. The predictions in either model are
in good agreement with experimental data for most of the cases
displayed above, especially given the uncertainties in the matrix
elements (see text). It may be noted that values of
$S(B_d\rightarrow J/\psi K_S)$ as high as 0.74 in the AB model,
and as low as 0.65 in the BPW model, can be achieved by varying
the fit.} \vspace*{5pt}

In obtaining the entries for the $K$-system we have used central
values of the matrix element $\hat{B}_K$ and the loop functions
$\eta_i$ (see Refs. \cite{Buras, Ciuchinietal} for definitions and
values) characterizing short distance QCD effects  - i.e.
$\hat{B}_K = 0.86\pm 0.13,\ f_K = 159\ MeV,\ \eta_1 = 1.38 \pm
0.20$, $\eta_2 = 0.57 \pm 0.01$ and $\eta_3 = 0.47 \pm 0.04$. For
the $B$-system we use the central values of the unquenched lattice
results: $f_{B_d}\sqrt{\hat{B}_{B_d}} = 215(11)(^{+0}_{-23})(15)\
MeV$ and $f_{B_s}\sqrt{\hat{B}_{B_s}} =
245(10)(^{+3}_{-2})(^{+7}_{-0})\ MeV$ \cite{SAoki}. Note that the
uncertainties in some of these hadronic parameters are in the
range of 15\%; thus the predictions of the two SO(10) models as
well as those of the SM would be uncertain at present to the same
extent.

Some points of distinctions and similarities between the two
models are listed below.

(1) First note that the data point $(m_o,\ m_{1/2})$ = (300, 300)
GeV displayed above, though consistent with CP violation, gives
too large a value for Br($\mu\to e\gamma$) for both BPW and AB
models. All other cases shown in table 4 are consistent with the
experimental limit on $\mu\to e\gamma$ for the BPW model. For the
AB model on the other hand, as may be inferred from table 1, the
choice $(m_o,\ m_{1/2})$ = (1000, 1000) GeV is the only case that
is consistent with the limit on $\mu\to e\gamma$ (see table 1). It
is to be noted that for this case the squark masses are extremely
high ($\sim 2.8$ TeV), and therefore, in the AB model, once the
$\mu\to e\gamma$ constraint is satisfied, the SUSY contributions
are strongly suppressed for all four entities: $\Delta m_K,\
\epsilon_K,\ \Delta m_{B_d}$ and $S(B_d\rightarrow J/\psi K_S)$.

(2) For the BPW model on the other hand, there are good regions of
parameter space allowed by the limit on the rate of $\mu\to
e\gamma$ (e.g. $(m_o,~m_{1/2})=(600,~300)$ GeV), which are also in
accord with $\epsilon_K$. The SUSY contribution to $\epsilon_K$
for these cases is sizable ($\sim 20-30\%$) and negative, as
desired.

(3) We have exhibited the case $(m_o,\ m_{1/2})$ = (1000, 250) GeV
to illustrate that this case does not work for either model as it
gives too low a value for $\epsilon_K$ in the BPW model, and a
negative value in the AB model. In this case the SUSY
contribution, which is negative, is sizable because of the
associated loop functions which are increasing functions of
$(m_{sq}^2/m_{\tilde{g}}^2)$.

(4) The predictions regarding $\Delta m_K$, $\Delta m_{B_d}$ and
$S(B_d\rightarrow J/\psi K_S)$ are very similar in both the
models, i.e they are both close to the SM value.

(5) As noted above, there are differences between the predictions
of the BPW vs. the AB models for $\epsilon_K$ for a given $(m_o,\
m_{1/2})$. With uncertainties in $\hat{B}_K$ and the SUSY
spectrum, $\epsilon_K$ cannot, however, be used at present to
choose between the two models, but if $(m_o,\ m_{1/2})$ get
determined (e.g. following SUSY searches at the LHC) and
$\hat{B}_K$ is more precisely known through improved lattice
calculations, $\epsilon_K$ can indeed distinguish between the BPW
and the AB models, as also between SO(10) and G(224) models (for
details on this see Ref. \cite{BPR}). This distinction can be
sharpened especially by searches for $\mu\to e\gamma$.

(6) {\bf $B_d\rightarrow \phi K_S,\ \Delta m_{B_s}$}: Including
the SM$'$ and SUSY contributions to the decay $B_d\rightarrow \phi
K_S$, we get the following results for the CP violating asymmetry
parameter S($B_d\rightarrow \phi K_S$) in the two models:
\begin{eqnarray}
\begin{array}{l}
{\rm\bf BPW:\ \ }S(B_d\rightarrow\phi K_S)\approx
+0.65-0.74~.\\
\begin{array}{l}
{\rm\bf \ \ AB:\ \ }S(B_d\rightarrow\phi K_S)\approx
+0.61-0.65~.\\
\end{array}
\end{array}
\end{eqnarray}
\noindent The values displayed above for the AB model are
calculated for the fit given in Eq. (\ref{eq:fitAB}). For variant
fits in the AB model, values as high as S($B_d\rightarrow \phi
K_S)\approx$ 0.7 may be obtained. The SUSY contribution to the
amplitude for the decay $B_d\rightarrow \phi K_S$ in the BPW model
is only of order 1\%, whereas in the AB model it is nearly 5\% for
light SUSY spectrum ($(m_o,\ m_{1/2})\sim(300,300)$ GeV) and about
1\% for large $(m_o,\ m_{1/2})(\sim(1000,500)$ GeV). The main
point to note is that in both models S($B_d\rightarrow \phi K_S$)
is positive in sign and close to the SM prediction. The current
experimental values for the asymmetry parameter are
$S(B_d\rightarrow\phi K_S)=(+0.50\pm
0.25^{+0.07}_{-0.04})_{BaBar}; (+0.06\pm 0.33\pm 0.09)_{BELLE}$
\cite{BabarBelleNew}\footnote{At the time of completing this
manuscript, the BELLE group reported a new value of
$S(B_d\rightarrow\phi K_S)=+0.44\pm 0.27^\pm 0.05$ at the 2005
Lepton-Photon Symposium \cite{NewBelle}. This value is close to
the value reported by BaBar, and enhances the prospect of it being
close to the SM prediction.}. While the central values of these
two measurements are very different, the errors on them are large.
It will thus be extremely interesting from the viewpoint of the
two frameworks presented here to see whether the true value of
$S(B_d\rightarrow\phi K_S)$ will turn out to be close to the
SM-prediction or not.

Including SUSY contributions to $B_s-\overline{B_s}$ mixing coming
from $\delta^{23}_{LL,RR,LR,RL}$ insertions we get:
\begin{eqnarray}
\begin{array}{l}
{\bf BPW:\ \ }\Delta m_{B_s}(Tot\approx SM')\approx {\bf 17.3}\
ps^{-1} \bigl(\frac{f_{B_s}\sqrt{\hat{B}_{B_s}}}{245
MeV}\bigr)^2~.\\
\begin{array}{l}
{\bf \ \ AB:\ \ }\Delta m_{B_s}(Tot\approx SM')\approx {\bf 16.6}\
ps^{-1} \bigl(\frac{f_{B_s}\sqrt{\hat{B}_{B_s}}}{245
MeV}\bigr)^2~.\\
\end{array}
\end{array}
\end{eqnarray}

\noindent Both predictions are compatible with the present lower
limit on $\Delta m_{B_s}\gsim 14.4 ps^{-1}$ \cite{ASoni}.

(7) {\bf Contribution of the A term to $\epsilon'_K$}: Direct CP
violation in $K_L\rightarrow\pi\pi$ receives a new contribution
from the chromomagnetic operator $Q^-_g = (g/16\pi^2) (\bar s_L
\sigma^{\mu\nu} t^a d_R - \bar s_R \sigma^{\mu\nu} t^a
d_L)G^a_{\mu\nu}$, which is induced by the gluino penguin diagram.
This contribution is proportional to
$Im[(\delta^d_{LR})_{21}-(\delta^d_{LR})_{12}^{*}]$, which is
known in both models (see Eqs. (\ref{eq:AdBPW}) and
(\ref{eq:AdAB})). Following Refs. \cite{Buras2} and \cite{Nir},
one obtains:
\begin{eqnarray}
\label{eq:epsilon'} Re(\epsilon'/\epsilon)_{\tilde{g}}\approx 91\
B_G \bigl(\frac{110~ MeV}{m_s + m_d}\bigr) \bigl(\frac{500~
GeV}{m_{\tilde{g}}}\bigr)~
Im[(\delta^d_{LR})_{21}-(\delta^d_{LR})_{12}^{*}]
\end{eqnarray}
where $B_G$ is the relevant hadronic matrix element.
Model-dependent considerations (allowing for $m_K^2/m_{\pi}^2$
corrections) indicate that $B_G \approx  1-4$, and that it is
positive \cite{Buras2}. Putting in the values of
$\delta^d_{LR})_{12,21}$ obtained in each model with ($m_o,\
m_{1/2}$) = (a) (600, 300) GeV, and (b) (1000, 1000) GeV, we get:
\begin{eqnarray}
\begin{array}{l}
{\bf BPW:\ \ \ } Re(\epsilon'/\epsilon)_{\tilde{g}}\approx
+(3.7\times 10^{-4})(B_G/4)(10/\tan\beta)\quad Case\ (a)~.\\
\begin{array}{l}
~~~~~~~~~~~~~~~~~~~~~~~~ \approx
+(4.5\times 10^{-5})(B_G/4)(10/\tan\beta)\quad Case\ (b)~.\\
\begin{array}{l}
{\bf \ \ AB:\ \ } Re(\epsilon'/\epsilon)_{\tilde{g}}\approx
-(3.7\times 10^{-5})(B_G/4)(10/\tan\beta)\quad Case\ (a)~.\\
\begin{array}{l}
~~~~~~~~~~~~~~~~~~~~~\approx
+(4.5\times 10^{-6})(B_G/4)(10/\tan\beta)\quad Case\ (b)~.\\
\end{array}
\end{array}
\end{array}
\end{array}
\end{eqnarray}
Whereas both cases (a) and (b) are consistent with the limit on
$\mu\to e\gamma$ for the BPW model, only case (b) is in accord
with $\mu\to e\gamma$ for the AB model. The observed value of
$Re(\epsilon'/\epsilon)_{obs}$ is given by
$Re(\epsilon'/\epsilon)_{obs} = (17\pm 2)\times 10^{-4}$
\cite{AAlavi}. At present the theoretical status of SM
contribution to $Re(\epsilon'/\epsilon)$ is rather uncertain. For
instance, the results of Ref. \cite{TBlum} and \cite{Noaki} based
on quenched lattice calculations in the lowest order chiral
perturbation theory suggest negative central values for
$Re(\epsilon'/\epsilon)$. (To be specific Ref. \cite{TBlum} yields
$Re(\epsilon'/\epsilon)_{SM}=(-4.0\pm 2.3)\times 10^{-4}$, the
errors being statistical only.) On the other hand, using methods
of partial quenching \cite{Golterman} and staggered fermions,
positive values of $Re(\epsilon'/\epsilon)$ in the range of about
$(3-13)\times 10^{-4}$ are obtained in \cite{Bhattacharya}. In
addition, a recent non-lattice calculation based on
next-to-leading order chiral perturbation theory yields
$Re(\epsilon'/\epsilon)_{SM}=(19\pm 2^{+9}_{-6}\pm 6)\times
10^{-4}$ \cite{Pich}. The systematic errors in these calculations
are at present hard to estimate. The point to note here is that
the BPW model predicts a relatively large and positive SUSY
contribution to $Re(\epsilon'/\epsilon)$, especially for case (a),
which can eventually be relevant to a full understanding of the
value of $\epsilon'_K$, whereas this contribution in the AB model
is rather small for both cases. Better lattice calculations can
hopefully reveal whether a large contribution, as in the BPW
model, is required or not.

(8) {\bf EDM of the neutron and the electron}: RG-induced
$A$-terms of the model generate chirality-flipping sfermion mixing
terms $(\delta_{LR}^{d,u,l})_{ij}$, whose magnitudes {\it and}
phases are predictable in the two models (see Eq.
(\ref{eq:deltalr})), for a given choice of the universal
SUSY-parameters ($m_o,\ m_{1/2},\ and\ \tan\beta$). These
contribute to the EDM's of the quarks and the electron by
utilizing dominantly the gluino and the neutralino loops
respectively. We will approximate the latter by using the
bino-loop. These contributions are given by (see e.g.
\cite{Nath}):
\begin{eqnarray}
\begin{array}{c}
(d_d, d_u)_{A_{ind}} = (-\frac{2}{9},
\frac{4}{9})\frac{\alpha_s}{\pi}\ e\
\frac{m_{\tilde{g}}}{m_{sq}^2}
f\bigl(\frac{m^2_{\tilde{g}}}{m_{sq}^2}\bigr)
Im(\delta^{d,u}_{LR})_{11}\\
\begin{array}{cc}
(d_e)_{A_{ind}} = -\frac{1}{4\pi}\frac{\alpha_{em }}{cos^2
\theta_W}\ e\ \frac{m_{\tilde{B}}}{m_{\tilde{l}}^2}
f\bigl(\frac{m^2_{\tilde{B}}}{m_{\tilde{l}}^2}\bigr)
Im(\delta^l_{LR})_{11} ~.\\
\end{array}
\end{array}
\end{eqnarray}
The EDM of the neutron is given by $d_n = \frac{1}{3}(4d_d-d_u)$.
The up sector being purely real implies $d_u=0$ in the AB model.
In table 5 we give the values of $d_n$ and $d_e$ calculated in the
two models for moderate and heavy SUSY spectrum and $\tan\beta =
10$. \vspace*{12pt}

\begin{tabular}{|c|c|c|c|c|}
\hline \rule[-3mm]{0mm}{8mm} ($m_o,\ m_{1/2}$)(GeV) &
\multicolumn{2}{c|}{AB-model}&\multicolumn{2}{c|} {BPW-model}\\
\hline
& $d_n$ (e-cm)&  $d_e$ (e-cm) & $d_n$ (e-cm) &  $d_e$ (e-cm) \\

\hline

I (600, 300)& $4.0\times 10^{-26}$ & $1.6\times 10^{-27}$
         & 1.1$\times 10^{-26}$  & 1.1$\times10^{-29}$
                  \\ \hline
II (1000, 500)& $1.4\times 10^{-26}$ & $5.9\times 10^{-28}$
         & 3.9$\times 10^{-27}$  & 4.1$\times10^{-30}$
                  \\ \hline
III (1000, 1000)& $5.7\times 10^{-27}$ & $1.1\times 10^{-27}$
         & 1.7$\times 10^{-27}$  & 7.7$\times10^{-30}$
                  \\ \hline
Expt. upper bound& $6.3\times 10^{-26}$ &$4.3\times
10^{-27}$&$6.3\times 10^{-26}$&$4.3\times 10^{-27}$
 \\ \hline
\end{tabular}
\vskip.10in \noindent Table 5. {\small EDMs of neutron and
electron calculated in the BPW and the AB models for moderate and
heavy SUSY spectrum and $\tan\beta = 10$ arising only from the
induced A-terms. While all cases are consistent with $\mu\to
e\gamma$ for the BPW model, only case III is consistent for the AB
model.} \vspace*{5pt}

From the table above, we see that while both models predict that
the EDM of the neutron should be seen within an improvement by a
factor of 5--10 in the current experimental limit, their
predictions regarding the EDM of the electron are quite different.
While the AB model predicts that the EDM of the electron should be
observed with an improvement by a factor of 5--10 in the current
experimental limit, the prediction of the BPW model for the EDM of
the electron is that it is 2 to 3 orders of magnitude smaller than
the current upper bound. These predictions are in an extremely
interesting range; while future experiments on edm of the neutron
can provide support for or deny both models, those on the edm of
the electron can clearly distinguish between the two models.

\section{Conclusions}

In this paper we did a comparative study of two realistic SO(10)
models: the hierarchical Babu-Pati-Wilczek (BPW) model and the
lop-sided Albright-Barr (AB) model. Both models have been shown to
successfully describe fermion masses, CKM mixings and neutrino
oscillations. Here we compared the two models with respect to
their predictions regarding CP and flavor violations in the quark
and lepton sectors. CP violation is assumed to arise primarily
through phases in fermion mass matrices (see e.g. Ref.
\cite{BPR}). For all processes we include the SM as well as SUSY
contributions. For the SUSY contributions, assuming that the SUSY
messenger scale $M^*$ lies above $M_{GUT}$ as in a mSUGRA model,
we include contributions from both post-GUT physics as well as
those arising due to RG running in MSSM below the GUT scale. While
this has been done before for the BPW model in Refs. \cite{BPR}
and \cite{LFV}, this is the first time that flavor and CP
violations have been studied in the AB model {\it including} both
post-GUT and sub-GUT physics. This inclusion brings out important
distinctions between the two models.

Previous works on lepton flavor violation in the AB model
\cite{ABLFV} have included only the RHN contribution associated
with sub-GUT physics. It is important to note, however, that in
both models the sfermion-transition elements
$\delta^{ij}_{LL,RR,LR,RL}$ and the induced A parameters get fully
determined for a given choice of soft SUSY-breaking parameters
($m_o,\ m_{1/2},\ A_o,\ \tan\beta$ and $sgn(\mu)$) and thus both
contributions are well determined. Including both contributions,
we find the following similarities and distinctions between the
two models.

\noindent {\bf Similarities:}\\
$\bullet$ Both models are capable of yielding values of the
Wolfenstein parameters ($\rho'_W,\ \eta'_W$) which are close to
the SM values and simultaneously the right gross pattern for
fermion masses, CKM elements and neutrino oscillations. For this
reason, both models give the values of $\Delta m_K,\ \Delta
m_{B_d}$ and $S(B_d\to J/\psi K_S)$ that are close to the SM
predictions and agree quite well with the
data. The SUSY contribution to these processes is small ($\lsim 3\% $). \\
$\bullet$ For the case of $\epsilon_K$, it is found that for the
BPW model, the SM$'$ value is larger than the observed value by
about 20\% for central values of $\hat{B}_K$ and $\eta_i$, but the
SUSY contribution is sizable and negative, so that the net value
can be in good agreement with the observed value for most of the
SUSY parameter space. For the AB model, for the choice of input
parameters as in Eq. (\ref{eq:fitAB}), the SM$'$ value for
$\epsilon_K$ is close to the observed value.  For most of the
soft-SUSY parameter space the AB model also yields $\epsilon_K$ in
good agreement with the observed value once one allows for
uncertainties in the matrix elements (see table 4).
\\ $\bullet$ Both models predict that $S(B_d\to\phi K_S)$ should be
$\approx +0.65-0.74$, close to the SM predictions. \\ $\bullet$
The predictions regarding $\Delta m_{B_s}$ are similar and
compatible with the experimental limit in both models.\\ $\bullet$
Both models predict the EDM of the neutron to be ($few \times
10^{-26} e-cm$) which should be observed with an improvement in
the current limit by a factor of 5--10.

Thus a confirmation of these predictions on the edm of the neutron
and $S(B_d\to\phi K_S)$, would go well with the two models, but
cannot distinguish between them.

\noindent {\bf Distinctions:}\\
$\bullet$The lepton sector brings in impressive distinction
between the two models through lepton flavor violation and through
the EDM of the electron as noted below.\\ $\bullet$ The BPW model
gives BR($\mu\to e\gamma$) in the range of $10^{-11}-10^{-13}$ for
slepton masses $\lsim 500$ GeV with the restriction that
$m_{1/2}\lsim 300$ GeV (see remarks below table 1). Thus it
predicts that $\mu\to e\gamma$ should be seen in upcoming
experiments which will have a sensitivity of
$10^{-13}-10^{-14}$\cite{mueupcoming}. The contribution to $\mu\to
e\gamma$ in the AB model is generically much larger than that of
the BPW model. For it to be consistent with the experimental upper
bound on BR($\mu\to e\gamma$), the AB model would require a rather
heavy SUSY spectrum, i.e. $(m_o,\ m_{1/2})\gsim (1000,\ 1000)$
GeV, i.e. $m_{\tilde{l}}\gsim 1200$ GeV and $m_{\tilde{q}}\gsim
2.8$ TeV. With the constraints on $(m_o,\ m_{1/2})$ as noted
above, both models predict that $\mu\to e\gamma$ should be seen
with an improvement in the current limit which needs to be a
factor of 10--50 for the BPW model and a factor of 3--5 for the AB
model.

$\bullet$ An interesting distinction between the AB and the BPW
models arises in their predictions for the EDM of the electron.
The AB model give $d_e$ in the range $10^{-27}-10^{-28} e\ cm$
which is only a factor of 3--10 lower than the current limit. Thus
the AB model predicts that the EDM of the electron should be seen
in forthcoming experiments. The BPW model on the other hand
predicts a value of $d_e$ in the range $10^{-29}-10^{-30} e\ cm$
which is about 100--1000 times lower than the current limit.

$\bullet$ In the quark sector, another interesting distinction
between the two models comes from $\epsilon'/\epsilon$. The BPW
model predicts that Re($\epsilon'/\epsilon$)$_{SUSY}\approx
+5\times 10^{-4} (B_G/4)(10/\tan\beta)$. Thus the BPW model
predicts that SUSY will give rise to a significant positive
contribution to $\epsilon'/\epsilon$, assuming $B_G$ is positive
\cite{Buras2}. The AB model gives
Re($\epsilon'/\epsilon$)$_{SUSY}\approx -5\times
10^{-5}(B_G/4)(10/\tan\beta)$. Thus it predicts that the SUSY
contribution is $\sim\mathcal{O}(1/10)$ the experimental value and
is negative. Since the current theoretical status of the SM
contribution to Re($\epsilon'/\epsilon$) is uncertain, the
relevance of these contributions can be assessed only after the
associated matrix elements are known reliably.

In conclusion, the Babu-Pati-Wilczek model and the Albright-Barr
model have both been extremely successful in describing fermion
masses and mixings and neutrino oscillations. In this note,
including all three important sources of flavor violation (two of
which have been neglected in the past), we have seen that CP and
flavor violation can bring out important distinctions between the
two models, especially through studies of $\mu\to e\gamma$ and the
edm of the electron. It will be extremely interesting to see how
these two models fare against the upcoming experiments on CP and
flavor violation.

\section{Acknowledgements}
I would like to thank Prof. Jogesh C. Pati for guidance and
helpful discussions, and Prof. Kaladi S. Babu for many useful
insights and suggestions. I would also like to thank Profs. Carl
H. Albright and Stephen M. Barr for taking the time to go through
this manuscript and for their comments.
\newpage
\section{Figures}
\begin{figure}[!ht]
\centering
\includegraphics[scale=5,height=3in,width=5in]{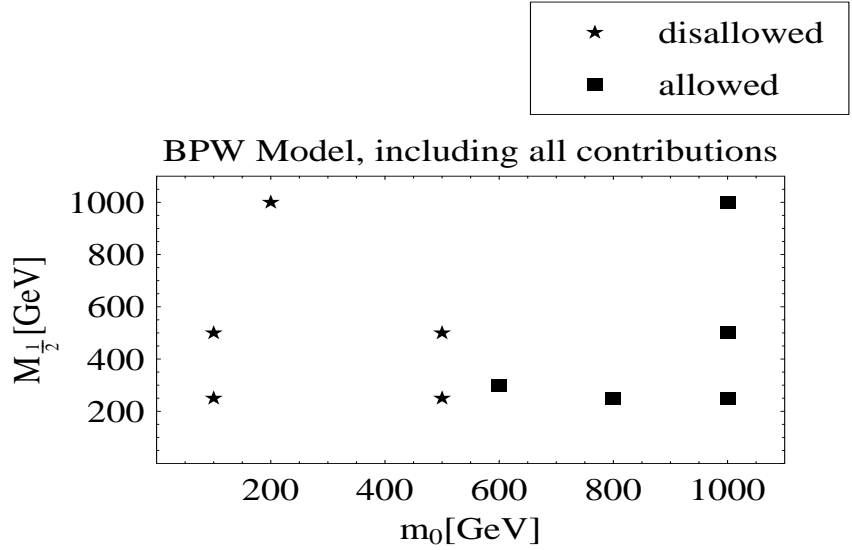}
\caption{Regions in the $(m_o,~m_{1/2})$ plane allowed and
disallowed by the current experimental limit on Br($\mu\to
e\gamma$)= $1.2\times 10^{-11}$ as obtained for the BPW model with
$\ln(M^*/M_{GUT})=1$, $\tan\beta=10$ and $\mu>0$. The points
allowed by the limit on Br($\mu\to e\gamma$) are marked with a
box, while the points disallowed by this limit are marked with a
star. The results include post-GUT and RHN contributions to the
rate of $\mu\to e\gamma$. Note that a large region of parameter
space is allowed.}
\end{figure}
\newpage
\begin{figure}[!ht]
\centering
\includegraphics[scale=5,height=3in,width=5in]{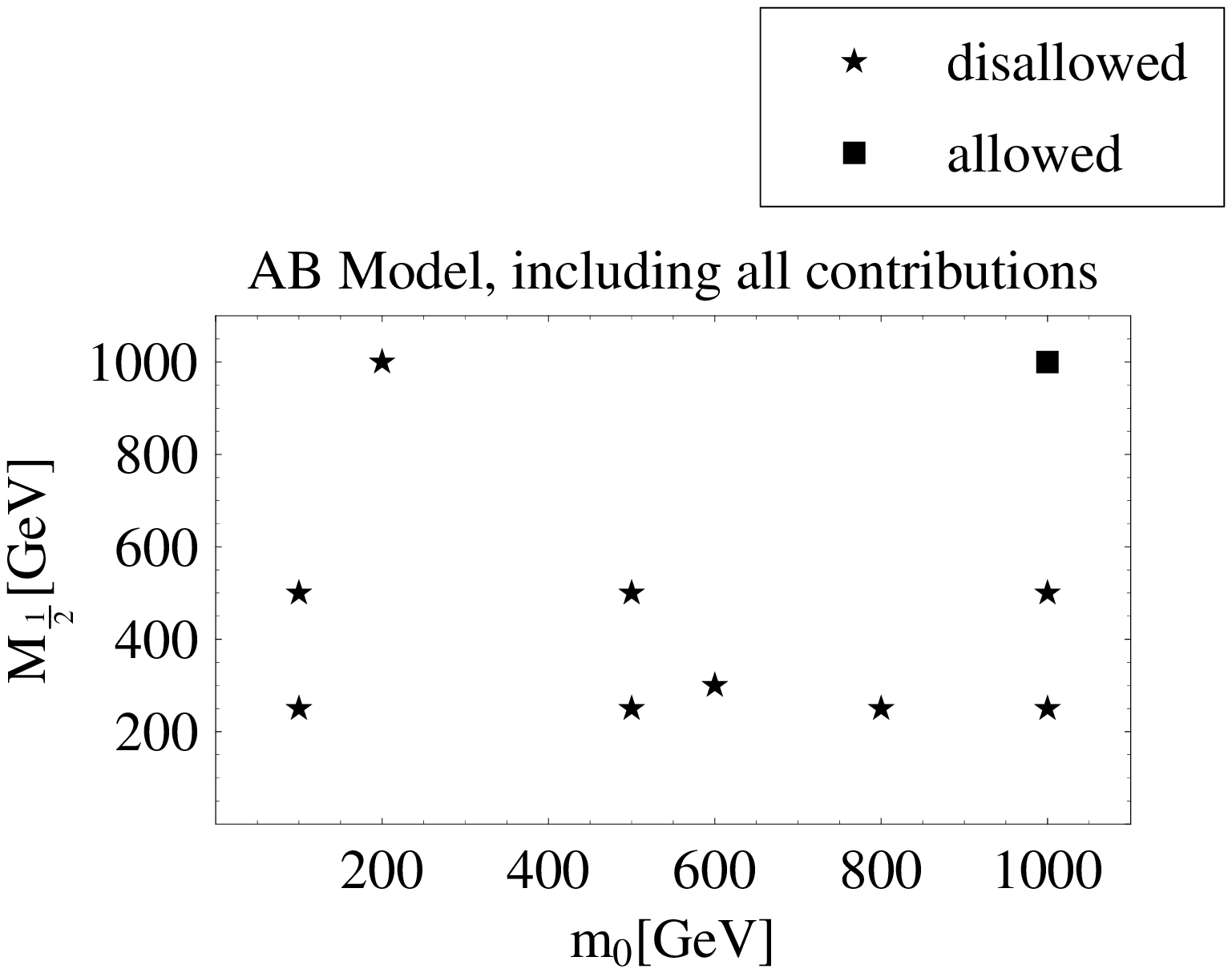}
\caption{Regions in the $(m_o,~m_{1/2})$ plane allowed and
disallowed by the current experimental limit on Br($\mu\to
e\gamma$)= $1.2\times 10^{-11}$ as obtained for the AB model with
$\ln(M^*/M_{GUT})=1$, $\tan\beta=10$ and $\mu>0$. The points
allowed by the limit on Br($\mu\to e\gamma$) are marked with a
box, while the points disallowed by this limit are marked with a
star.  The results include post-GUT and RHN contributions to the
rate of $\mu\to e\gamma$. Note that, only a rather heavy SUSY
spectrum with $(m_o,~m_{1/2})\gsim (1000,~1000)$ GeV is allowed by
the limit on $\mu\to e\gamma$. This corresponds to a squark mass
of $\sim 2.8$ TeV and a slepton mass of $\sim 1200$ GeV.}
\end{figure}
\begin{figure}[!ht]
\includegraphics[scale=1,height=3in,width=3in]{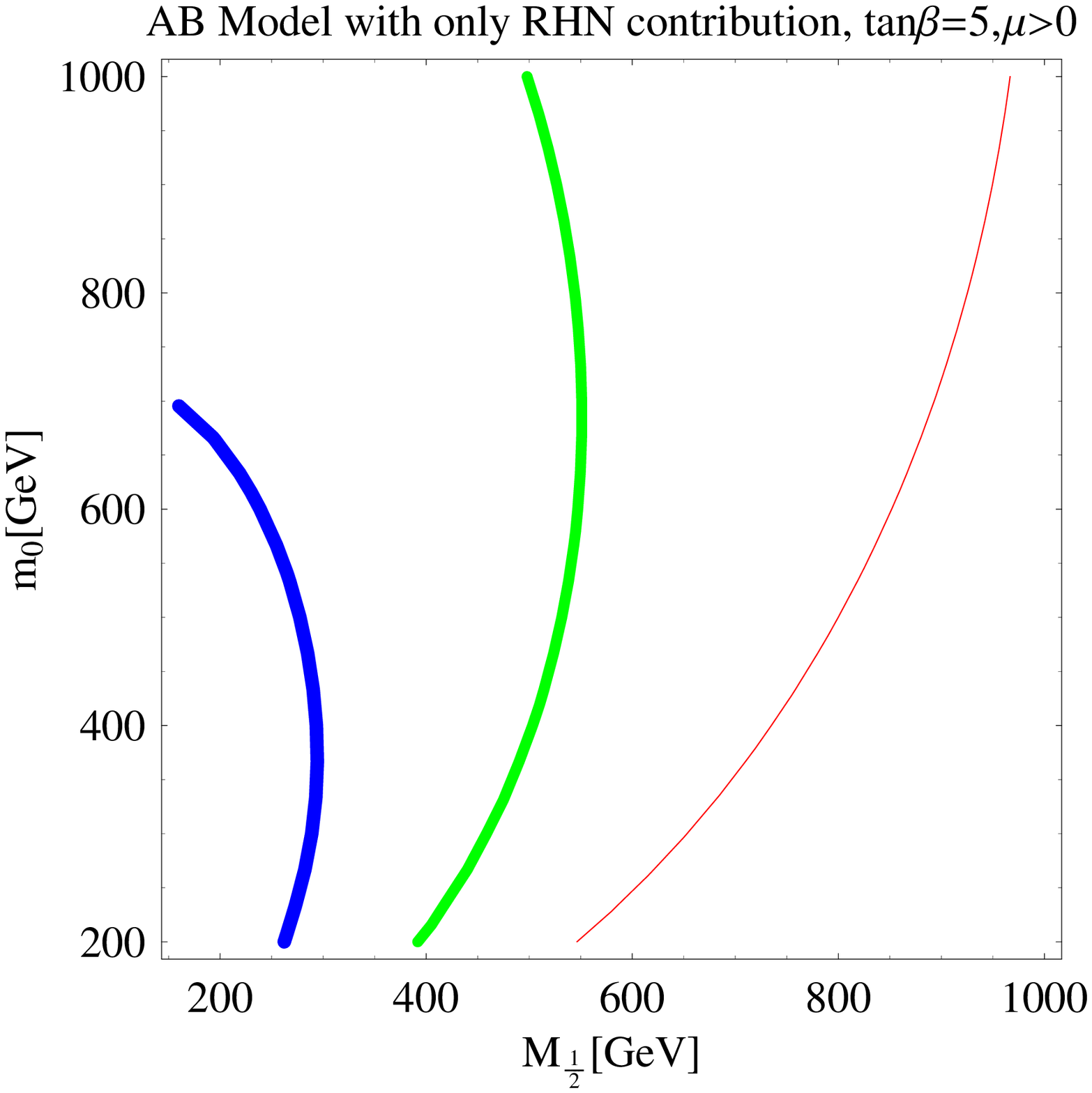}
\includegraphics[scale=1,height=3in,width=3in]{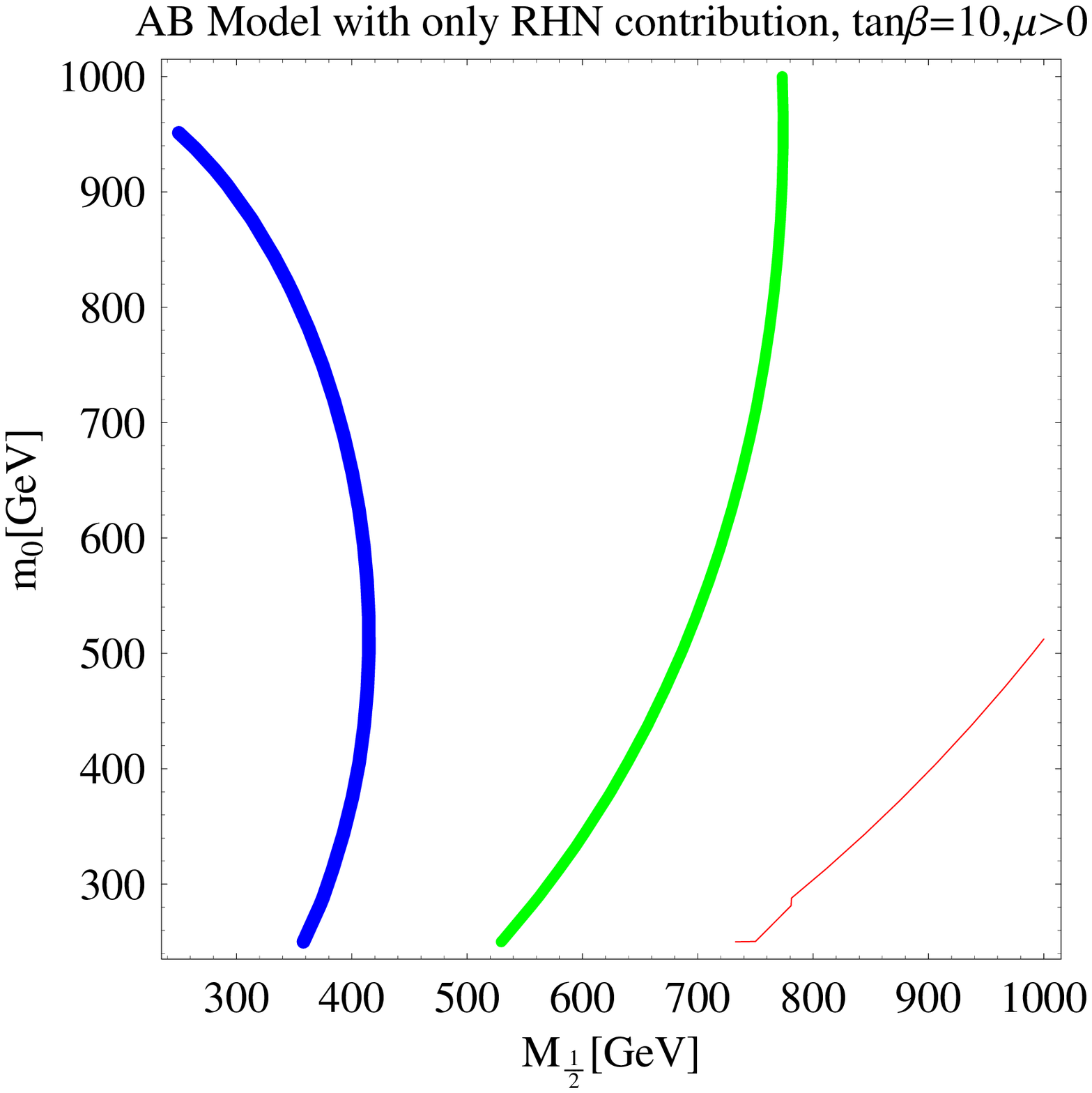}
\caption{Curves of constant Br($\mu\to e\gamma$) in the
$(m_o,~m_{1/2})$ plane with only the right handed neutrino
contribution for the case of the AB model. The thickest (blue)
line corresponds to the experimental limit of $1.2\times
10^{-11}$, the medium (green) line to Br($\mu\to
e\gamma)=10^{-12}$, and the thinnest (red) one to Br($\mu\to
e\gamma)=10^{-13}$. A similar analysis was carried out in Ref.
\cite{Jankowski}.}
\end{figure}

\end{document}